\documentclass[a4paper,11pt]{article}

\usepackage{moreverb}
\usepackage{graphicx,amssymb,amstext,amsmath,tikz}
\usepackage[para]{threeparttable}
\usetikzlibrary{positioning,shapes.geometric}
\usepackage{rotating}
\usepackage[T1]{fontenc}
\usepackage{longtable}

\usepackage{float}
\usepackage[caption = false]{subfig}
\usepackage{natbib}
\newcommand\ie{\textsl{i.e.}\ }

\usepackage[left=2cm,top=2cm,right=2cm, bottom=2cm]{geometry}
\def\s{\sigma^2}

\def\m{\mu}
\def\m1{\mu_1}
\def\m0{\mu_0}
\def\logit{\mbox{logit}}

\def\ci{_{1i}}

\def\Yc{Y_{1i}}
\def\Ye{Y_{2i}}
\def\ci{\perp\!\!\!\perp}

\date{} 
\begin{document}
\title{Methods for estimating complier-average causal effects for cost-effectiveness analysis}

\date{}
\author{K.~DiazOrdaz,  A.~J.~Franchini, R.~Grieve\\
Department of Medical Statistics,\\
 London School of Hygiene and Tropical Medicine,\\
\texttt{karla.diaz-ordaz@lshtm.ac.uk}}

\maketitle

\begin{abstract}

In Randomised Controlled Trials (RCT) with treatment non-compliance, instrumental variable approaches are used to estimate complier average causal effects. We extend these approaches to cost-effectiveness analyses, where methods need to recognise the correlation between cost and health outcomes. 

We propose a Bayesian full likelihood (BFL) approach, which jointly models the effects of random assignment on treatment received and the outcomes, and a three-stage least squares (3sls)
method, which acknowledges the correlation between the endpoints, and the endogeneity of the treatment received.

 This investigation is motivated by the REFLUX study, which exemplifies the setting where compliance differs between the RCT and routine practice. A simulation is used to compare the methods performance. 

We find that failure to model the correlation between the outcomes and treatment received
correctly can result in poor CI coverage and biased estimates. By contrast, BFL and 3sls methods provide unbiased estimates with good coverage.
\vskip 2ex
\scriptsize{\textbf{Keywords}: non-compliance, instrumental variables, bivariate outcomes, cost-effectiveness}

\end{abstract}

\maketitle

\section{Introduction}

 Non-compliance is a common problem in Randomised Controlled Trials (RCTs), as some participants depart from their randomised treatment, by for example switching from the experimental to the control regimen. An unbiased estimate of the effectiveness of treatment assignment can be obtained by reporting the intention-to-treat (ITT) estimand. In the presence of non-compliance, a complimentary estimand of interest is the causal effect of treatment received. Instrumental variable (IV) methods can be used to obtain the complier average causal effect (CACE), as long as random assignment meets the IV criteria for identification \citep{Angrist1996}. An established approach to IV estimation is two-stage least squares (2sls), which provides consistent estimates of the CACE when the outcome measure is continuous, and non-compliance is binary \citep{Baiocchi2014}.

Cost-effectiveness analyses (CEA) are an important source of evidence for informing clinical decision-making and health policy. CEA commonly report an ITT estimand, \ie the relative cost-effectiveness of the intention to receive the intervention \citep{NICE}. However, policy-makers may require additional estimands, such as the relative cost-effectiveness for compliers. For example, CEAs of new therapies for end-stage cancer, are required to estimate the cost-effectiveness of treatment receipt, recognising that patients may switch from their randomised allocation following disease progression.  Alternative estimates such as the CACE may also be useful when levels of compliance
in the RCT differ to those in the target population, or where intervention receipt, rather than the
intention to receive the intervention, is the principal cost driver. 
Methods for obtaining the CACE for univariate survival outcomes have been exemplified before \citep{Latimer2014}, but approaches for obtaining estimates that adequately adjust for non-adherence in CEA more generally, have received little attention. This has been recently identified as a key area where methodological development is needed \citep{Hughes2016}. 

The context of trial-based CEA highlights an important complexity that arises with multivariate outcomes more widely, in that, to provide accurate measures of the uncertainty surrounding a composite measure of interest, for example the incremental net monetary benefit (INB), it is necessary to recognise the correlation between the endpoints, in this case, cost and health outcomes \citep{Willan2003,Willan2006a}.   Indeed, when faced with non-compliance, and the requirement to estimate a causal effect of treatment on cost-effectiveness endpoints, some CEA resort to per protocol (PP) analyses \citep{Brilleman2015}, which exclude participants who deviate from treatment. As non-compliance is likely to be associated with prognostic variables, only some of which are observed, PP analyses are liable to provide biased estimates of the causal effect of the treatment received.

This paper develops novel methods for estimating CACE in CEA that use data from RCTs with
non-compliance. First, we propose using the three stage least squares (3sls) method \citep{Zellner1963}, which allows the estimation of a system of simultaneous equations with endogenous
regressors. Next, we consider a bivariate version of the `unadjusted Bayesian' models previously
proposed for Mendelian randomisation \citep{Burgess2012}, which simultaneously estimate
the expected treatment received as a function of random allocation, and the mean outcomes as a linear
function of the expected treatment received.  Finally, we develop a Bayesian full likelihood approach (BFL), whereby the outcome variables and the treatment received are jointly modelled as dependent on the random assignment. This is an extension to the multivariate case of what is known in the econometrics literature as the IV unrestricted reduced form \citep{Kleibergen2003}.

The aim of this paper is to present and compare these alternative approaches. The problem is illustrated in Section \ref{Sec:Reflux} with the REFLUX study, a multicentre RCT and CEA that contrasts laparoscopic surgery with medical management for patients with Gastro-Oesophageal Reflux Disease (GORD). Section \ref{Sec:CACE} introduces the assumptions and methods for estimating CACEs. Section \ref{Sec:sim}  presents a simulation study used to assess the performance of the alternative approaches, which  are then  applied to the case study in Section \ref{Sec:RefluxCACE}. We conclude with a Discussion (Section \ref{Sec:Discussion}), where we consider the findings from this study in the context of related research.

\section{Motivating example: Cost-effectiveness analysis of the REFLUX study}\label{Sec:Reflux}
The REFLUX study was a UK multicentre RCT with a parallel design, in which patients with moderately severe GORD, were randomly assigned to medical management or laparoscopic surgery  \citep{Grant2008, Grant2013}.

 The RCT randomised 357  participants (178 surgical, 179 medical) from 21 UK centres. An observational preference based study was conducted alongside it, which involved 453 preference participants (261 surgical, 192 medical). 


For the cost-effectiveness analysis within the trial,  individual resource use (costs in $\pounds$ sterling) and health-related quality of life (HRQoL), measured using EQ5D (3 levels), were recorded annually for up to 5 years. The HRQoL data were used to adjust life years and present quality-adjusted life years (QALYs) over the follow-up period \citep{Grant2013}.\footnote{There was no administrative censoring.}  As is typical, the costs were right-skewed. Table 1 reports the main characteristics of the data set.

 The original CEA estimated the linear additive treatment effect on mean costs and health outcomes (QALYs). The primary analysis used a system of seemingly unrelated regression equations (SURs) \citep{Zellner1962, Willan2004}, adjusting for baseline HRQoL EQ5D summary score (denoted by $\mbox{EQ5D}_{0}$). The SURs can be written for  cost $Y_{1i}$ and QALYs $Y_{2i}$, as follows
\begin{equation}\label{sur}
\begin{array}{c}
Y_{1i} = \beta_{0,1}+ \beta_{1,1} \mbox{treat}_i + \beta_{1,2} \mbox{$\mbox{EQ5D}_{0}$}_{i} + \epsilon_{1i}\\
Y_{2i}= \beta_{0,2}+\beta_{1,2} \mbox{treat}_i + \beta_{2,2} \mbox{$\mbox{EQ5D}_{0}$}_{i} + \epsilon_{2i}
\end{array}
\end{equation}
where $\beta_{1,1}$ and $\beta_{1,2}$ represent the incremental costs and QALYs respectively.  The error terms are required to satisfy
$E[\epsilon_{1i}]=E[\epsilon_{2i}]=0, \ 
E[\epsilon_{ki} \epsilon_{k'i}] =\sigma_{kk'},\ 
E[\epsilon_{ki} \epsilon_{k'j}]= 0,\ \mbox{\ for\ }k,\ k'\in\{1,2\}, \mbox{\ and for\ }i\neq j.$ 
Rather than assuming that the errors are drawn from a bivariate normal distribution,  estimation is usually done by the feasible generalized least squares (FGLS) method\footnote{If we are prepared to assume the errors are bivariate normal, estimation can proceed by maximum likelihood.}. This is a two-step method where, in the first step, we run ordinary least squares estimation for equation \eqref{sur}. In the second step, residuals from the first step are used as estimates of the elements $\sigma_{kk'}$ of the covariance matrix, and this estimated covariance structure is then used to re-estimate the coefficients in equation \eqref{sur}  \citep{Zellner1962,Zellner1962a}.

In addition to reporting incremental costs and  QALYs, CEA often report the incremental cost-effectiveness ratio (ICER), which is defined as the ratio of the incremental costs per incremental QALY, and the incremental net benefit (INB), defined as
$\mbox{INB}(\lambda) = \lambda \beta_{1,2}-\beta_{1,1}$, where  $\lambda$ represents the decision-makers' {\it willingness to pay} for a one unit gain in health outcome. Thus the new treatment is cost-effective if $\mbox{INB}>0$. For a given $\lambda$, the standard error of INB can be estimated from the estimated increments $\hat{\beta}_{1,1}$ and $\hat{\beta}_{1,2}$, together with their standard errors and their correlation following the usual rules for the variance of a linear combination of two random variables.
The willingness to pay $\lambda$  generally lies in a range, so it is common to compute the estimated value of INB for various values of $\lambda$. In REFLUX, the reported INB was calculated using  $\lambda= \pounds 30000$, which is within the range of cost-effectiveness thresholds used by the UK National Institute for Health and Care Excellence \citep{NICE}. 

The original ITT analysis concluded that, compared to medical management, the arm assigned to surgery had a large gain in average QALYs, at a small additional cost and was relatively cost-effective with a positive mean INB, albeit with 95\% confidence intervals (CI) that included zero. However, these ITT results cannot  be interpreted as a causal effect of the treatment, since within one year of randomisation, 47 of those randomised to surgery switched and received medical management, while in the medical treatment arm, 10 received surgery. The reported reasons for not having the allocated surgery were that in the opinion of the surgeon or the patient, the symptons were not ``sufficiently severe'' or the patient was judged unfit for surgery (e.g. overweight).
The preference-based observational study conducted alongside the RCT reported that in routine clinical practice, the corresponding proportion who switched from an intention to have surgery and received medical management, was relatively low (4\%), with a further 2\% switching from control to intervention \citep{Grant2013}.
Since the percentage of patients who switched in the RCT was higher than in the target population
and the costs of the receipt of surgery are relatively large, there was interest in reporting a causal
estimate of the intervention. Thus, the original study also reported a PP analysis on complete-cases, adjusted for baseline $\mbox{EQ5D}_{0}$, which resulted in an ICER of $\pounds 7263$ per additional QALY \citep{Grant2013}.
This is not an unbiased estimate of the causal treatment effect, so in Section  \ref{Sec:RefluxCACE}, we re-analyse the REFLUX dataset  to obtain a CACE of the cost-effectiveness outcomes, recognising the joint distribution of costs and QALYs, using the methods described in the next section.

\section{Complier Average Causal effects with bivariate outcomes}\label{Sec:CACE}

We begin by defining more formally our estimands and assumptions. Let $\Yc$ and $\Ye$ be the continuous bivariate outcomes, and $Z_i$ and $D_i$ the binary random treatment allocation and treatment received respectively, corresponding to the $i$-th individual. The bivariate endpoints $Y_{1i}$ and $Y_{2i}$ belong to the same individual $i$, and thus are correlated. We assume that there is an unobserved confounder $U$, which is associated with the treatment received and either or both of the outcomes.
From now on, we will assume that the \textbf{(i) Stable Unit Treatment Value Assumption (SUTVA)} holds:
the potential outcomes of the $i$-th individual are unrelated to the treatment status of all other individuals (known as \textit{no interference}), and that for those who actually received treatment level $z$, their observed outcome is the potential outcome corresponding to that level of treatment.

Under SUTVA, we can write the potential  treatment received by the $i$-th subject  under the  random assignment at level $z_i \in \{0, 1\}$ as  $D_i\left(z_i\right)$. Similarly, $Y_{\ell i}\left(z_i,d_i\right)$ with $\ell\in\{1,2\}$ denotes the corresponding potential outcome for endpoint $\ell$, if the $i$-th subject were allocated to level $z_i$ of the treatment and received level $d_i$. There are four potential outcomes. Since each subject is randomised to one level of treatment, only one of the potential outcomes per endpoint $\ell$, is observed, \ie $Y_{\ell i}=Y_{\ell i}(z_i,D_i(z_i))=Y_i(z_i)$.

The CACE for outcome $\ell$ can now be defined as
\begin{equation}\label{cace}
\theta_{\ell}=E\left[\{Y_{\ell i}(1)-Y_{\ell i}(0)\}\big|\{D_i(1)-D_i(0)=1\}\right].
\end{equation}

In addition to (i) SUTVA,  the following  assumptions are sufficient for identification of the CACE, \citep{Angrist1996}: \vskip 1ex
\noindent\textbf{(ii) Ignorability of the treatment assignment}: $Z_i$ is independent of unmeasured confounders (conditional on measured covariates)  and the potential outcomes $Z_{i} \ci U_i,D_{i}(0),D_{i}(1),Y_{i}(0),Y_{i}(1).$
\\ \textbf{(iii) The random assignment predicts treatment received}: $Pr\{D_i(1)=1\}\neq Pr\{D_i(0)=1\}$.
\\  \textbf{(iv) Exclusion restriction}: The effect of $Z$ on $Y_\ell$ must be via an effect of $Z$ on $D$; $Z$ cannot affect $Y_\ell$ directly. \\ \textbf{(v) Monotonicity:} $D_i(1)\geq D_i(0)$.

 The CACE can now be identified from equation \eqref{cace} without any further assumptions about the unobserved confounder; in fact, $U$ can be an effect modifier of the relationship of $D$ and $Y$ \citep{Didilez2010}.

In the REFLUX study, the assumptions concerning the random assignment, (ii) and (iii), are justified by design. The exclusion restriction assumption seems plausible for the costs, since the costs of surgery are only incurred if the patient actually has the procedure. We argue that it is also plausible that it holds for QALYs, as the participants did not seem to have a preference for either treatment, thus making the psychological effects of knowing to which treatment one has been allocated minimal. The monotonicity assumption rules out the presence of defiers.  It seems fair to assume that there are no participants who would refuse the REFLUX surgery when randomised to it, but who would receive surgery when randomised to receive medical management.   Equation \eqref{cace} implicitly assumes that receiving the intervention has the same average effect in the linear scale, regardless of the level of $Z$ and $U$.  This average is however across different `versions' of the intervention, as  the trial protocol did not prescribe a single surgical procedure, but allowed for the surgeon to choose their preferred laparoscopy method, as would be the case in routine clinical practice.

Since random allocation, $Z$, satisfies assumptions (ii)-(iv), we say it is an instrument (or instrumental variable) for $D$. For binary instrument, the simplest method of estimation of equation \eqref{cace} in the IV framework is the Wald estimator \citep{Angrist1996}:
\[
\hat{\theta}_{\ell, IV} = \frac{E(Y|Z=1) - E(Y|Z=0)}{E(D|Z=1) - E(D|Z=0)}
\]
Typically, estimation of these conditional expectations proceeds via an approach known as two-stage least squares (2sls). The first stage fits a linear regression to treatment received on treatment assigned. Then, in a second stage,  a regression model for the outcome on the predicted treatment received is fitted:
\begin{eqnarray}\label{2sls}
\nonumber D_i &= &\alpha_0 + \alpha_1Z_i + \omega_{1i}\\
Y_{\ell i} &=& \beta_0 + \beta_{IV} \hat{D_i} + \omega_{2i}
\end{eqnarray}
\noindent where $\hat{\beta}_{IV} $ is an estimator for $\theta_{\ell}$. Covariates can be used, by including them in both stages of the model. To obtain the correct standard errors for the 2sls estimator, it is necessary to take into account the uncertainty about the first stage estimates. The asymptotic standard error for the 2sls CACE is given in \citet{Imbens1994}, and implemented in commonly used software packages.

OLS estimation  produces first-stage residuals $\omega_{1i}$ that are uncorrelated with the instrument, and this is sufficient to guarantee that the 2sls estimator is consistent for the CACE \citep{AngristBook}. Therefore, 
we restrict our attention here to models where  the first-stage equation is linear, even though the treatment received is binary. \footnote{Non-linear versions of the 2sls exist. See for example \citet{Clarke2012} for an excellent review of methods for binary outcomes.} 

A key issue for settings such as CEA where there is interest in estimating the CACE for bivariate outcomes, is that 2sls as implemented in most software packages can only be  readily applied to univariate outcomes. 
Ignoring the correlation between the two endpoints is a concern for obtaining standard errors of composite measures of the outcomes, e.g. INB, as this requires accurate estimates of the covariance between the outcomes of interest (e.g. costs and QALYs).

A simple way to address this problem would be to apply 2sls directly to the composite measure, i.e. a net-benefit two-stage regression approach \citep{Hoch2006}. However, it is known that net benefit regression is very sensitive to outliers, and to distributional assumptions \citep{Willan2004}, and has been recently shown to perform poorly when these assumptions are thought to be violated \citep{Mantopoulos2016}. Moreover, such net benefit regression is restrictive, in that it does not allow separate covariate adjustment for each of the component outcomes, (e.g. baseline HRQoL, for the QALYs as opposed to the costs). In addition, this simple approach would not be valid for estimating the ICER, which is a non-linear function of the  incremental costs and QALYs. For these reasons, we do not consider this approach further. Rather, we present here three flexible strategies for estimating a CACE of the QALYs and the costs, jointly. The first approach combines SURs (equation \ref{sur}) and 2sls (equation \ref{2sls}) to obtain CACEs for both outcomes accounting for their correlation. This simple approach is known in the econometrics literature as three-stage least squares (3sls).

\subsection{Three-stage least squares}
Three-stage least squares (3sls) was developed for SUR systems with {\it endogenous} regressors, \ie any explanatory variables which are correlated with the error term in equations \eqref{sur} \citep{Zellner1963}. All the parameters appearing in the system are estimated jointly, in three stages. The first two stages are as for 2sls, but with the second stage applied to each of the outcomes. 
\begin{eqnarray}\label{3sls}
\nonumber\mbox{(1st stage):}\qquad D_i &=& \alpha_0 + \alpha_1Z_i + e_{0i}\\
\mbox{(2nd stage):}\qquad Y_{1i} &=& \beta_{01} + \beta_{IV,1} \hat{D_i} + e_{1i} \\
\label{3sls2}Y_{2i} &=& \beta_{02} + \beta_{IV,2} \hat{D_i} + e_{2i}
\end{eqnarray}
As with 2sls, the models can be extended to include baseline covariates. 
The third stage is the same step used on a SUR with exogenous regressors (equation \eqref{sur}) for estimating the covariance matrix of the error terms from the two equations \eqref{3sls} and \eqref{3sls2}. Thus, because we are assuming that $Z$ satisfies the identification assumptions (i)-(v),
Z is independent of the residuals at first and second stage, i.e.  $Z \ci e_{0i}$, $Z \ci e_{1i}$, and $Z \ci e_{2i}$.  Then, the 3sls procedure allows us to obtain the covariance matrix between the residuals $e_{1i}$ and $e_{2i}$. As with SURs, the 3sls approach does not require to make any distributional assumptions, as estimation can be done  by FGLS, and it is robust to heteroscedasticity of the errors in the linear models for the outcomes \citep{Greene2002}. 
  We note that the 3sls estimator based on FGLS is consistent only if the error terms in each equation of the system and the instrument are independent, which is likely to hold here, as we are dealing with a randomised instrument. In settings where this condition is not satisfied, other estimation approaches such as generalised methods of moments (GMM) warrant consideration   \citep{Schmidt1990}. In the just-identified case, \ie when there are as many endogenous regressors as there are instruments,  classical theory about 3sls estimators shows that the GMM and the FGLS estimators coincide \citep{Greene2002}. As the 3sls method uses an estimated variance-covariance matrix, it is only asymptotically efficient \citep{Greene2002}.

\subsection{Naive Bayesian estimators}
  Bayesian models have a natural appeal for cost-effectiveness analyses, as they afford us the flexibility to estimate bivariate models on the expectations of the two outcomes using different distributions, as proposed by \citep{Nixon2005}.
These models are often specified by writing a marginal model for one of the outcomes, e.g. the costs $Y_1$, and then, a model for $Y_2$, conditional on $Y_1$.

For simplicity of exposition, we begin by assuming normality for both outcomes and no adjustment for covariates. We have a marginal model for $Y_1$ and a model for $Y_2$  conditional on $Y_1$ \citep{Nixon2005}
\begin{eqnarray}\label{eq:qaly}
\label{eq:cost}
 Y_{1i}&\sim& N(\mu_{1i}, \s_1)\qquad\qquad\quad\  \mu_{1i} = \beta_{0,1}+ \beta_{1,1} \mbox{treat}_i\\
 Y_{2i}\mid Y_{1i}&\sim& N(\mu_{2i}, \s_2(1-\rho^2)) \qquad
\mu_{2i}= \beta_{0,2}+\beta_{1,2}\mbox{treat}_i+ \beta_{2,2} (y_{1i}-\mu_{1i}),
\end{eqnarray}
where $\rho$ is the correlation between the outcomes.  The linear relationship between the two outcomes is represented by $\beta_{2,2}=\rho\frac{ \sigma_2}{\sigma_1}$.

Because of the non-compliance, in order to obtain a causal estimate of treatment, we need to add a linear model for the treatment received, dependent on randomisation $Z_i$, similar to the first equation of a 2sls. Formally,  this model (denoted uBN, for unadjusted Bayesian Normal) can be written with three equations as follows:
\begin{equation}\label{uBN}
\begin{array}{l}
D_{i} \sim N(\mu_{0i}, \s_0) \\
 Y_{1i}\sim N(\mu_{1i}, \s_1)\\
 Y_{2i}\big| Y_{1i}\sim N(\mu_{2i}, \s_2(1-\rho^2)) \\
\end{array}\ \ \ \ \ \ \ \
\begin{array}{l}
\mu_{0i} = \beta_{0,0} + \beta_{1,0}Z_i \\
\mu_{1i} = \beta_{0,1}+ \beta_{1,1} \mu_{0i}  \\
\mu_{2i} =  \beta_{0,2}+\beta_{1,2} \mu_{0i}+ \beta_{2,2} (y_{1i}-\mu_{1i})
\end{array}
\end{equation}
 This model is a bivariate version of the `unadjusted Bayesian' method previously proposed for Mendelian randomisation \citep{Burgess2012}. It is called unadjusted, because the variance structure of the outcomes is assumed to be independent of the treatment received. The causal treatment effect for outcome $Y_\ell$, with $\ell\in\{1,2\}$, is represented by  $\beta_{1,\ell}$ in equations \eqref{uBN}.
We use the Fisher's z-transform of $\rho$, i.e. $z={1 \over 2}\log \left({1+\rho \over 1-\rho}\right)$, for which we assume a vague normal prior, i.e. $z\sim N(0, 10^2)$.  We also use vague multivariate normal priors for the regression coefficient (with a precision of 0.01). For standard deviations, we use $\sigma_j\sim\mbox{Unif}(0,10)$, for $j \in\{0,1,2\}$. This is similar to the priors used in \citep{Lancaster2004},  and are independent of the regression coefficient of treatment received on treatment allocation $\beta_{1,0}$. 

Cost data are notoriously right-skewed, and Gamma-distributions are often used to model them. Thus, we can relax the normality assumption of equation \eqref{uBN}, and model $Y_1$ (\ie cost) with a Gamma distribution, and treatment received (binary), with a logistic regression. The health outcomes, $Y_2$, are still modelled with a normal distribution, as is customary. Because we are using a non-linear model for the treatment received, we use the predicted raw residuals from this model as extra regressors in the outcome models, similar to the 2-stage residual inclusion estimator \citep{Terza2008}. We model  $Y_1$ by its marginal distribution (Gamma) and  $Y_2$ by a conditional Normal distribution, given $Y_1$ \citep{Nixon2005}. We call this model `unadjusted Bayesian Gamma-Normal' (uBGN)  and write it as follows
\begin{eqnarray}\label{equBGN}
\begin{array}{l}
 D_i \sim \mbox{Bern}(\pi_{i}) \\
 Y_{1i}\sim \mbox{Gamma}(\nu_{1i}, \kappa_1)\\
Y_{2i}\mid Y_{1i}\sim N(\mu_{2i}, \s_2(1-\rho^2)) \\
\end{array}\ \ \ \ \ \ \ \
\begin{array}{l}
 \logit(\pi_{i})  = \alpha_0 + \alpha_1Z_i \\
 r_{i} = D_i-\pi_i \\
\mu_{1i}= \beta_{0,1}+ \beta_{1,1} D_{i}+ \beta_{1,r} r_{i} \\
\mu_{2i}= \beta_{0,2}+\beta_{1,2}D_{i}+ \beta_{2,r} r_{i} +\beta_{2,2} (y_{1i}-\mu_{1i})
\end{array}
\end{eqnarray}
where $\mu_{1}=\frac{\nu_1}{\kappa_1}$, is the mean of the Gamma distributed costs, with shape $\nu_1$ and rate $\kappa_1$.
Again, we express  $\beta_{2,2}= \rho\frac{ \sigma_2 }{\sigma_1}$, and   assume a vague Normal prior  on the Fisher's z-transform of  $\rho$,  $z\sim N(0, 10^2)$.
The prior distribution for $\nu_1$ is $\mbox{Gamma}(0.01, 0.01)$. We also assume a Gamma prior for the intercept term of the cost equation, $\beta_{0,1}\sim\mbox{Gamma}(0.01, 0.01)$.  All the other regression parameters have the same priors as those used in the uBN model.

The models introduced in this section, uBN and uBGN, are estimated in one stage, allowing feedback between the regression equations and the propagation of uncertainty.  However, these 'unadjusted' methods ignore the correlation between the outcomes and the treatment received. This misspecification of the covariance structure may result in biases in the causal effect, which are likely to be more important at higher levels of non-compliance.

\subsection{Bayesian simultaneous equations (BFL)}

We now introduce an approach that models the covariance between treatment received and outcomes appropriately, using a  system of simultaneous equations. This can be done via full or limited information maximum likelihood, or by using MCMC to estimate the parameters in the model simultaneously allowing for proper Bayesian feedback and propagation of uncertainty. Here, we propose a Bayesian approach which is an extension of the methods presented in \citet{Burgess2012, Kleibergen2003, Lancaster2004}.

This method treats the endogenous variable $D$ and the cost-effectiveness outcomes as covariant and estimates the effect of treatment allocation, as follows. Let $(D_{i}, Y_{1i}, Y_{2i})^\top$ be the transpose of vector of outcomes, which now includes treatment received, as well as the bivariate endpoints of interest.
We treat all three variables as multivariate normally distributed, so that
 \begin{equation}
\begin{array}{l}
\left(\!\!\begin{array}{c}
D_{i}\\ Y_{1i} \\ Y_{2i}\end{array}\!\!\right)
\sim N\left\{\!\!\left(\begin{array}{c}
\mu_{0i}\\
\mu_{1i}\\
\mu_{2i}\end{array}\right),  \Sigma= \left(\!\!\!\begin{array}{ccc} \s_{0} & s_{01} & s_{02} \\  s_{01}& \s_{1} & s_{12} \\
 s_{02}& s_{12} & \s_{2}
 \end{array} \!\!\!\right)
\!\!\right\};
\end{array}\ \ \
\begin{array}{l}
\mu_{0i} =\beta_{0,0} + \beta_{1,0}Z_i \\
\mu_{1i} = \beta_{0,1}+ \beta_{1,1} \beta_{1,0}Z_i  \\
\mu_{2i}=  \beta_{0,2}+\beta_{1,2}\beta_{1,0}Z_i
\end{array}
 \end{equation}
where $s_{ij}=\mbox{cov}(Y_i,Y_j)$, and the causal treatment effect estimates are $\beta_{1,1}$ and $\beta_{1,2}$ respectively.  For the implementation, we use vague normal priors for the regression coefficients, \ie $\beta_{m,j}\sim N(0, 10^2)$, for $j \in\{0,1,2\}, m\in\{0,1\}$, and a Wishart prior for the inverse of $\Sigma$ \citep{GelmanHill}.

\section{Simulation study}\label{Sec:sim}

 We now use a factorial simulation study to assess the finite sample performance of the alternative methods. The first factor is the proportion of participants who do not comply with the experimental regime, when assigned to it, expressed as a percentage of the total (one-sided non-compliance). Bias is expected to increase with increasing levels of non-compliance.  A systematic review \citep{Dodd2012}  found that the percentage of non-compliance was less than 30\% in two-thirds of published RCTs, but greater than 50\% in one-tenth of studies. Here, two levels of non-compliance are chosen,  30\% and 70\%. As costs are typically skewed, three different distributions (Normal, Gamma or Inverse Gaussian -- IG) are used to simulate cost data. As the 2sls approach fails to accommodate the correlation between the endpoints, we examined the impact of different levels of correlation on the methods' performance; $\rho$ takes one of four values $\pm 0.4,\ \pm 0.8$. The final factor is the sample size of the RCT, taking two settings $n= 100, \text{and}\  1000$. In total, there are $2\times 3\times 4 \times 2=48$ simulated scenarios.

To generate the data, we begin by simulating  $U\sim N(0.50, 0.25^2)$,  independently from treatment allocation. $U$ represents a pre-randomisation variable that is a common cause of both the outcomes and the probability of non-compliance, \ie it is a confounding variable, which we assume is unobserved.

Now, let $S_i\sim \mbox{Bern}(\pi_s)$ be the random variable denoting whether the $i$th individual switches from allocated active treatment to control. The probability $\pi_s$ of one-way non-compliance with allocated treatment depends on $U$, in the following way,
\begin{equation}
\pi_s=\begin{cases} p+0.1, \text{if\ } u>0.5,\\
p-0.1, \text{otherwise}
\end{cases}
\end{equation}
where $p$ denotes the corresponding average non-compliance percentage expressed as a probability, i.e here $p \in \{0.3, 0.7\}$.
We now generate $D_i$, the random variable of treatment received as
\begin{equation}
D_i=\begin{cases} Z_i, \text{if either\ } s_i=0\text{\  or\ } Z_i=0,\\
1-Z_i, \text{if\ } s_i=1 \text{ and } Z_i=1
\end{cases}
\end{equation}
where $Z_i$ denotes the random allocation for subject $i$.

Then, the means for both outcomes are assumed to depend linearly on treatment received and the unobserved confounder $U$ as follows:
\begin{eqnarray}\label{meancost}\mu_1= E[Y_1]&=&1.2+ 0.4 D_i + 0.16(u_i - 0.5)\\
\mu_2=E[Y_2]&=&0.5+0.2 D_i + 0.04(u_i - 0.5)
\end{eqnarray}

Finally, the bivariate outcomes are generated using Gaussian copulas, initially with normal marginals. In subsequent scenarios, we consider Gamma or Inverse Gaussian marginals for $Y_1$ and normal for $Y_2$. The conditional correlation between the outcomes, $\rho$, is set according to the corresponding scenario.

For the scenarios where the endpoints are assumed to follow a bivariate normal distribution, the variances of the outcomes are set to $\s_1=0.2^2, \s_2=0.1^2$ respectively, while for scenarios with Gamma and IG distributed $Y_1$, the shape parameter is $\eta=4$. For the Gamma case, this  gives a variance for $Y_1$ equal to $\s_1=0.36$ in control and $\s_1=0.64$ in the intervention group.  When $Y_1\sim \mbox{IG}(\mu_1,\eta)$, the expected variance in the control group is $\s_1=0.432$, and $\s_1=1.024$  in those receiving the intervention.

The simulated endpoints represent cost-effectiveness variables that have been rescaled for computational purposes, with costs divided by $1000$, and QALYs by $0.1$, such that the true values are $\pounds 400$ (incremental costs), and $0.02$ (incremental QALYs) and so with a threshold value of $\lambda=\pounds 30000$ per QALY,  the true causal INB is $\pounds 200$. 

For each simulated scenario, we obtained $M=2500$ sets.
For the Bayesian analyses, we  use the median of the posterior distribution as the `estimate' of the parameter of interest, and the standard deviation of the posterior distribution as the standard error. Equi-tailed 95\% posterior credible intervals are also obtained. We use the term confidence interval for the Bayesian credible intervals henceforth, to have a unified terminology for both Bayesian and frequentist intervals.

Once the corresponding causal estimate has been obtained in each of the 2500 replicated sets under each scenario in turn, we compute the median bias of the estimates, coverage of 95\% confidence intervals (CI), median CI width and root mean square error (RMSE).  We report median bias as opposed to mean bias, because the BFL leads to a posterior distribution of the causal parameters which is Cauchy-like \citep{Kleibergen2003}.
A method is `adequate', if it results in low levels of bias (median bias $\le5$\%) with coverage rates within 2.5\% of the nominal value.

\textit{Implementation:}\\
The 3sls was fitted using \texttt{systemfit} package in R using FGLS, and the Bayesian methods were run using JAGS from R \texttt{(r2jags)}. Two chains, each one with 5000 initial iterations and 1000 burn-in were used. The multiple chains allowed for a check of convergence by the degree of their mixing and the initial iterations enabled to estimate iteration autocorrelation. A variable number of further 1000-iteration runs were performed until convergence was reached as estimated by the absolute value of the Geweke statistics for the first 10\% and last 50\% of iterations in a run being below 2.5. A final additional run of 5000 iterations was performed for each chain to achieve a total sample of 10000 iterations, and a MC error of about 1\% of the parameter SE on which to base the posterior estimates.
For the uBGN,  an offset of  0.01 is added to the shape parameter $\nu_1$ for the Gamma distribution of the cost,  to prevent the sampled shape parameter to become too close to zero, which may result in infinite densities. 
See the Appendix for the JAGS model code for BFL.

\subsection{Simulation Study Results}
\begin{figure}
\begin{center}
\caption{\footnotesize{Median Bias for scenarios with  30\% non-compliance and sample sizes of (a) $n=100$ (top) and (b)  $n=1000$. Results are stratified by cost distribution, and correlation between cost and QALYs. The dotted line represents zero bias. Results for 2sls (not plotted) are identical to those for 3sls; uBGN was not applied to Normal cost data.}}\label{30Biasplot}
\vskip -.3ex
\subfloat[n=100]{\includegraphics[scale=.45]{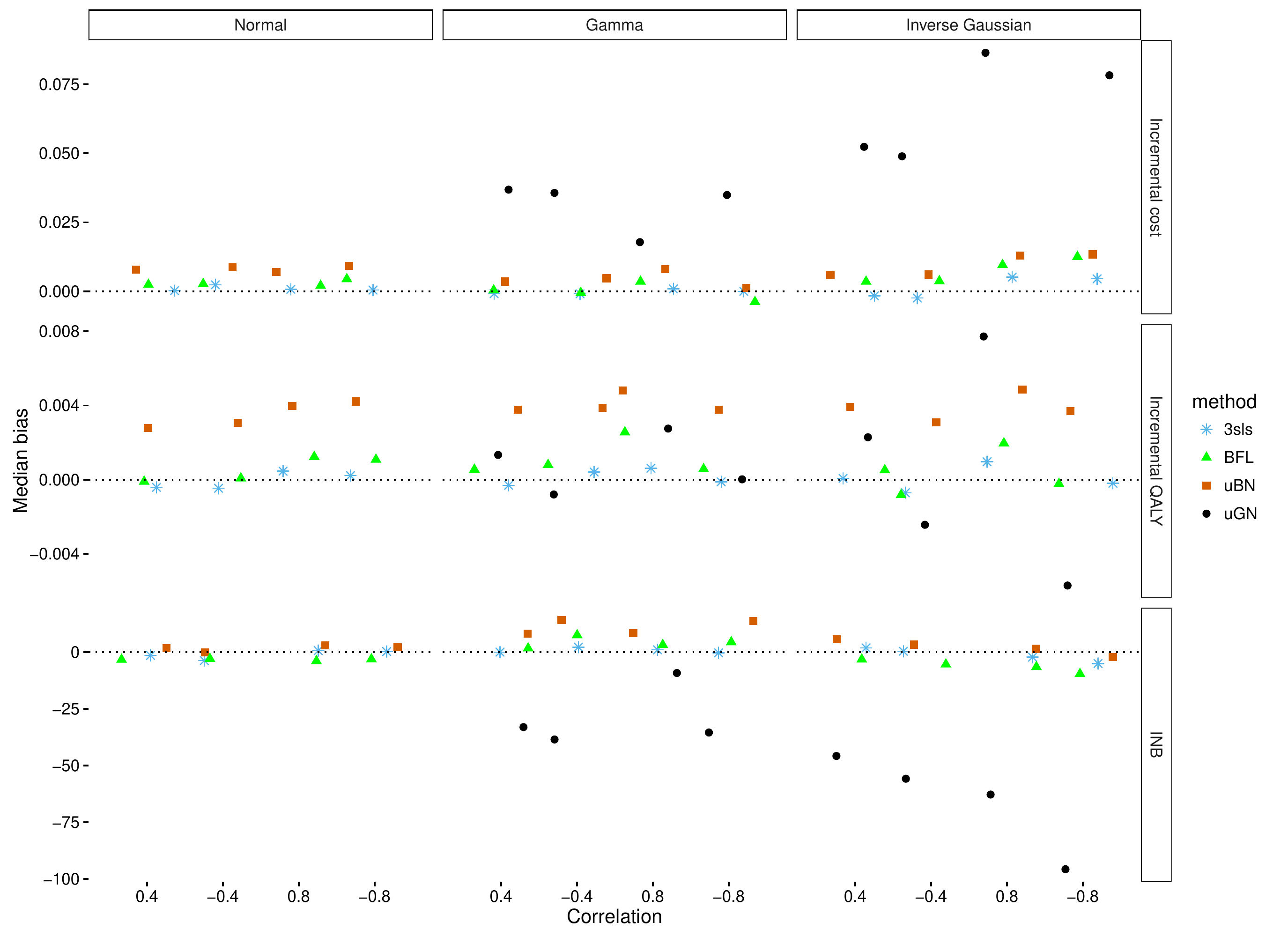}}\\\vskip -.45ex
\subfloat[n=1000]{\includegraphics[scale=.45]{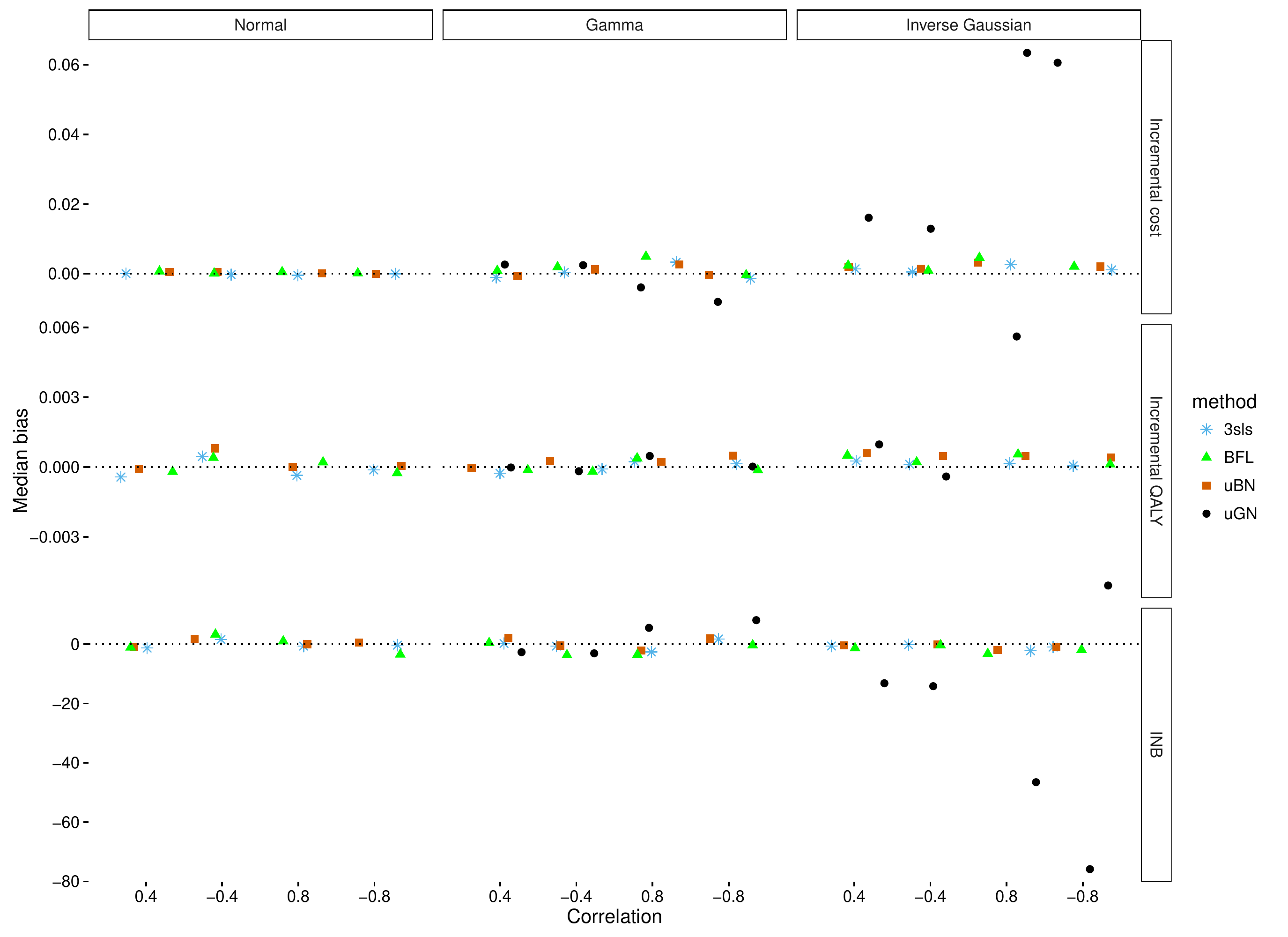}}
\vskip -.7ex
\end{center}
\end{figure}

\textbf{\textit{Bias}}\\
Figure \ref{30Biasplot} shows the median bias corresponding to scenarios with 30\% non-compliance, by cost distributions (left to right) and levels of correlation between the two outcomes, for sample sizes of $n=100$ (upper panel) and $n=1000$ (lower panel).
 With the larger sample size, for all methods, bias is negligible with normally distributed costs, and remain less than 5\% when costs are Gamma-distributed. However, when costs follow an Inverse Gaussian distribution, and the absolute levels of correlation between the endpoints are high ($\pm 0.8$), the uBGN approach results in biased estimates, around 10\% bias for the estimated incremental cost, and between 20 and 40\% for the estimated INB. With the small sample size and when costs follow a Gamma or Inverse Gaussian distribution, both unadjusted Bayesian methods provide estimates with moderate levels of bias.
With 70\% non-compliance (Figure A3  in the Appendix),  the unadjusted methods result in important biases which persist even with large sample sizes, especially for scenarios with non-normal outcomes. For small sample settings, uBN reports positive bias (10 to 20\%) in the estimation of incremental QALYs, and the resulting INB, irrespective of the cost distribution. The uBGN method reports relatively unbiased estimates of the QALYs, but large positive bias (up to 60\%) in the estimation of costs, and hence, there is substantial bias in the estimated INB (up to 200\%).  Unadjusted Bayesian methods estimates are biased in the direction of the true causal effect, because these methods ignore the positive correlation between treatment received and each outcome introduced by the confounding variable.  By contrast, the BFL and the 3sls provide estimates with low levels of bias across most settings.  
\\
\textbf{\textit{CI coverage and width}}\\
Table 2 presents the results for CI coverage and width, for scenarios with a sample size of $n=100$, absolute levels of correlation between the endpoints of 0.4, and 30\% non-compliance. All other results are shown in the Appendix.
The 2sls INB ignores the correlation between costs and QALYs, and thus, depending on the direction of this correlation, 2sls reports CI coverage that is above (positive correlation) or below (negative correlation) nominal levels. This divergence from nominal levels increases with higher absolute levels of correlation (see Appendix, Table A6).

The uBN approach results in over-coverage across many settings, with wide CIs. For example, for both levels of non-compliance and either sample size, when the costs are Normal, the CI coverage rates for both incremental costs and QALYs exceed 0.98.
The interpretation offered by  \citet{Burgess2012}  is also relevant here: the uBN assumes that the treatment received and the outcomes variance structures are uncorrelated, and so when the true correlation is positive, the model overstates the variance and leads to wide CIs.  By contrast, the uBGN method results in low CI coverage rates for the estimation of incremental costs, when costs follow an inverse Gaussian distribution. This is because the model incorrectly assumes a Gamma distribution, thereby underestimating the variance. The extent of the under-coverage appears to increase with higher absolute values of the correlation between the endpoints, with coverage as low as 0.68 (incremental costs) and 0.72 (INB)  in scenarios where the absolute value of correlation between costs and QALYs is 0.8. (see Appendix, Table A7).

The BFL approach reports estimates with CI coverage close to the nominal when the sample size is large, but with excess coverage (greater than 0.975), and relatively wide CI, when the sample size is $n=100$ (see Table 2 for 30\% noncompliance, and Table A4 in the Appendix for the result corresponding to 70\% non-compliance).  By contrast, the 3sls reports CI coverage within 2.5\% of nominal levels for each sample size, level of non-compliance, cost distribution and level of correlation between costs and QALYs.
\\
\textbf{\textit{RMSE}}\\
Table  2  reports RMSE corresponding to 30\% non-compliance, and $n=100$.  The least squares approaches result in lower RMSE than the other methods for the summary statistic of interest, the INB.  This pattern is repeated across other settings, see the Appendix, Tables A10--A16.

\section{Results for the motivating example}\label{Sec:RefluxCACE}

 We now compare the methods in practice by applying them to the REFLUX dataset. 
Only  48\%  of the individuals have completely observed cost-effectiveness outcomes: there were 185  individuals with missing  QALYs,  166 with missing costs, and a further   13 with missing $\mbox{EQ5D}_{0}$ at baseline, with about a third of those with missing outcomes having switched from their allocated treatment.  These missing data not only bring more uncertainty to our analysis, but more importantly, unless the missing data are handled appropriately can lead to biased causal estimates \citep{Daniel2011}.   A complete case analysis would be  unbiased, albeit inefficient, if  the missingness is conditionally independent of the outcomes given the covariates in the model  \citep{White2010}, even when the covariates have missing data, as is the case here.\footnote{This mechanism is a special case of missing not at random.} Alternatively, a more plausible assumption is to assume the missing data are missing at random (MAR), i.e. the probability of missingness depends only on the observed data, and use multiple imputation (MI) or a full Bayesian analysis to obtain valid inferences. 

Therefore, we perform MI prior to carrying out 2sls and 3sls analyses.   We begin by investigating  all the possible associations between the covariates available in the data set and the missingness, univariately for costs, QALYs and baseline $\mbox{EQ5D}_{0}$.   Covariates which are predictive  of both, the missing values  and the probability of  missing, are to be included in the imputation model as auxiliary variables, as conditioning on more variables helps make the MAR assumption more plausible.  None of the fully observed variables at our disposal (age, sex, and baseline BMI) satisfies these criteria and therefore, we do not include any auxiliary variables in our imputation models. Thus, we impute  total cost, total QALYs and baseline $\mbox{EQ5D}_{0}$, 50 times by  chained equations, using predictive mean matching  (PMM), taking the 5 nearest neighbours as donors  \citep{White2011}, including  treatment received in the imputation model and stratifying by treatment allocation.  
 We perform 2sls on costs and QALYs independently and calculate (within MI) SE for the INB assuming independence between costs and QALYs. For the 3sls approach, the model is fitted to both outcomes simultaneously, and  the post-estimation facilities are used to extract the variance-covariance estimate, and compute the estimated INB and its corresponding SE. We also use the CACE estimates of incremental cost and QALYs to obtain the ICER. After applying each method to the 50 MI sets, we combine the results using Rubin's rules \citep{Rubin1987}.\footnote{Applying IV 2sls and 3sls with multiply imputed datasets, and combining the results using Rubin's rules can be done automatically in Stata using \texttt{mi estimate, cmdok:  ivregress 2sls}  and \texttt{mi estimate, cmdok: reg3}. In R,  \texttt{ivregress} can be used with \texttt{with.mids} command within \texttt{mice}, but \texttt{systemfit} cannot presently be combined with this command so, Rubin's rules have to be coded manually. Sample code is available in the Appendix.}  

For the Bayesian approaches,  the missing values  become extra parameters to model. Since baseline $\mbox{EQ5D}_{0}$  has missing observations, a model for its distribution is added $\mbox{EQ5D}_{0}\sim N(\mu_{q0}, \sigma_{q0}^2)$, with  a vaguely informative prior for $\mu_{q0}\sim \mbox{Unif}(-0.5,1)$, and an uninformative  prior for $|\sigma_{q0}|\sim N(0, 0.01)$. 
We add two extra lines of code to the models to obtain posterior distributions for INB and ICERs. We center the outcomes around the empirical mean (except for costs, when modelled as Gamma) and re-scale the costs (dividing by 1000) to improve mixing and convergence. We use two chains, initially running 15,000 iterations with 5,000 as burn-in. After checking visually for auto-correlation, an extra 10,000 iterations are needed to ensure that the density plots of the parameters corresponding to the two chains are very similar, denoting convergence to the stationary distribution. Enough  iterations for each chain are kept to make the total effective sample (after accounting for auto-correlation) equal to 10,000. \footnote{Multivariate normal nodes cannot be partially observed in JAGS, thus,  we run BFL models on all available data within WinBUGs. An observation with zero costs  was set to missing when running the Bayesian Gamma model, which requires strictly positive costs.}

Table 1 shows the results for incremental costs, QALYs and INB for the motivating example adjusted for baseline $\mbox{EQ5D}_{0}$. Bayesian posterior distribution are summarised by their median value and 95\% credible intervals.  The CACEs are similar across the methods, except for uBGN, where the incremental QALYs CACE is nearly halved, resulting in a smaller INB with a CI that includes 0.   In line with the simulation results, this would suggest that, where the uBGN is misspecified according to the assumed cost distribution, it can provide a biased estimate of the incremental QALYs. 

Comparing the CACEs to the ITT, we see that the incremental cost estimates increases between an ITT and a CACE, as actual receipt of surgery carries with it higher costs that the mere offering of surgery does not. Similarly, the incremental QALYs are larger, meaning that amongst compliers, those receiving surgery have a greater gain in quality of life, over the follow-up period. The CACE for costs are relatively close to the per-protocol incremental costs reported in the original study, $\pounds 2324 \ (1780, 2848)$. In contrast, the incremental QALYs according to PP on complete-cases  originally reported was $ 0.3200\  (0.0837, 0.5562)$, considerably smaller than our CACE estimates \citep{Grant2013}. The ITT ICER obtained after MI was $\pounds 4135$, while using causal incremental costs and QALYs, the corresponding estimates of the CACE for ICER were  $\pounds 4140$ (3sls),  $\pounds 5189$ (uBN),  $\pounds   5960$ (uBGN), and $\pounds 3948$ (BFL). The originally reported per-protocol ICER is $\pounds 7932$ per extra QALY was obtained on complete cases only \citep{Grant2013}.

 These results may be sensitive  to the modelling of the missing data.  As a sensitivity analysis to the MAR assumption,  we present the complete case analysis in Table A1 in the Appendix.  The conclusions from complete-case analysis are  similar to those obtained under MAR.

We  also explore the sensitivity to choices of priors,  by  re-running the BFL analyses using different priors, first for the multivariate precision matrix, keeping the priors for the coefficients normal, and then a second analysis, with uniform priors for the regression coefficient, and an inverse Wishart prior with 6 degrees of freedom and a identity scale matrix, for the precision. The results are not materially changed (see Appendix, Table A2).

The results of  the within-trial CEA suggest that amongst compliers, laparoscopy is  more cost-effective than medical management for patients suffering from GORD.  The results are robust to the choice of priors, and to the assumptions about the missing data mechanism.   The results for the uBGN differ somewhat from the other models, and as our simulations show, the concern is that such unadjusted Bayesian models are prone to bias from model misspecification.  

\section{Discussion}\label{Sec:Discussion}

This paper extends existing methods for CEA \citep{Willan2003,Nixon2005}, by providing IV approaches for obtaining causal cost-effectiveness estimates for RCTs with non-compliance. The methods developed here however are applicable to other settings with multivariate continuous outcomes more generally, for example RCTs in education, with different measures of attainment being combined into an overall score. To help dissemination, we provide code in the Appendix.

We proposed exploiting existing 3sls methods and also considered IV Bayesian models, which are extensions of previously proposed approaches for univariate continuous outcomes. \cite{Burgess2012}  found the BFL was median unbiased and gave CI coverage close to nominal levels, albeit with wider CIs than least-squares methods. Their `unadjusted Bayesian' method, similar to our uBN approach, assumes that the error term for the model of treatment received on treatment  allocated is uncorrelated with the error from the outcome models. This results in bias in the direction of the true association, and affects the CI coverage. Our simulation study shows that, in a setting with multivariate outcomes, the bias can be substantial. A potential solution to this could be using priors  for the error terms that reflect this dependency explicitly. For example,  \citet{Rossi2012} propose using a prior for the errors that explicitly depends on the coefficient $\beta_{1,0}$, the effect of treatment allocation on treatment received, in equation \eqref{uBN}. \citet{Kleibergen2003} propose priors that also reflect this dependency explicitly, and replicate better the properties of the 2sls. This is known as the ``Bayesian two-stage approach''.

The results of our simulations show that applying 2sls separately to the univariate outcomes leads to inaccurate 95\% CI around the INB, even with moderate levels of correlation between costs and outcomes ($\pm 0.4$). Across all the settings considered, the 3sls approach resulted in low levels of bias for the INB and unlike 2sls, provided CI coverage close to nominal levels. BFL performed well with large sample sizes, but produced standard deviations which were too large when the sample size was small, as can be seen from the over-coverage, with wide CIs. 

The REFLUX study illustrated a common concern in CEA, in that the levels of non-compliance in the RCT were different, in this case higher, to those in routine practice. The CACEs presented provide the policy-maker with an estimate of what the relative cost-effectiveness would be if all the RCT participants had complied with their assigned treatment, which is complementary to the ITT estimate. Since we judged the IV assumptions for identification  likely to hold in this case-study, we conclude that either 3sls or BFL provide valid inferences for the CACE of INB. The re-analysis of the REFLUX case study also provided the opportunity to investigate the sensitivity to the choice of  priors in practice. Here we found that our choice of weakly informative priors, which were relatively flat in the region where the values of the parameters were anticipated to be, together with samples of at least size 100, had minimal influence on the posterior estimates. We repeated the analysis using different vague priors for the parameters of interests and the corresponding results were not materially changed.

We considered here relatively simple frequentist IV methods, namely 2sls and 3sls. One alternative approach to the estimation of CACEs for multivariate responses, is to use linear structural equation modelling, estimated by maximum-likelihood expectation-maximization  (ML-EM) algorithm \citep{Jo2001}. Further, we only considered those settings where a linear additive treatment effect is of interest, and the  assumptions for identification are met. Where interest lies in systems of simultaneous non-linear equations with endogeneous regressors, GMM or generalised  structural equation models can be used to estimate CACEs \citep{Davidson2004}.

There are several options to study the sensitivity to departures from the identification assumptions.  For example, if the exclusion restriction does not hold, a Bayesian parametric model can use priors on the non-zero direct effect of  randomisation on the outcome for identification \citep{Conley2012, Hirano2000}. Since the models are only weakly identified,  the results would depend strongly on the parametric choices for the  likelihood and the  prior distributions.  In the frequentist IV framework, such modelling is also possible, see \citet{Baiocchi2014} for an excellent tutorial on how to conduct sensitivity analysis to violations of the ER and monotonicity assumptions. 
Alternatively, violations of the ER can also be handled by using baseline covariates to model the probability of compliance directly, within structural equation modelling via ML-EM framework \citep{Jo2002a, Jo2002b}.

Settings where the instrument is only weakly correlated with the endogenous variable have not been considered here, because for binary non-compliance with binary allocation, the percentage of one-way non-compliance would need to be in excess of 85\%, for the F-statistic of the randomization instrument to be less than 10, the traditional cutoff beneath which an instrument is regarded as `weak'. Such levels of non-compliance are not realistic in practice, with the reported median non-compliance equal to 12\%  \citep{Zhang2014}. Nevertheless, Bayesian IV methods have been shown to perform better than 2sls methods when the instrument is weak  \citep{Burgess2012}.

 Also, for simplicity, we restricted our analysis of the case study to MAR and complete cases assumptions. Sensitivity to departures from these  assumptions is beyond the scope of this paper, but researchers should be aware of the need to think carefully about the possible causes of missingness, and  conduct sensitivity analysis under MNAR, assuming plausible differences in the distributions of the observed and the missing data.  When addressing the missing data through Bayesian methods, the posterior distribution can be sensitive to the choice of prior distribution,  especially  with a large amount of missing data \citep{Hirano2000}.

Future research directions could include exploiting the additional flexibility of the Bayesian framework to incorporate informative priors, perhaps as part of a comprehensive decision modelling approach. The methods developed here could also be extended to handle time-varying non-compliance.

{\scriptsize{
\subsection*{Acknowledgements}
We thank M. Sculpher, R. Faria, D. Epstein, C. Ramsey and the REFLUX study team for access to the data.\\
K.~DiazOrdaz was supported by UK  Medical Research Council Career development award in Biostatistics MR/L011964/1. This report is independent research supported by the National Institute for Health Research (Senior Research Fellowship, R.~Grieve, SRF-2013-06-016). The views expressed in this publication are those of the author(s) and not necessarily those of the NHS, the National Institute for Health Research or the Department of Health.}}

\begin{center}
\begin{table}
\caption{\footnotesize{The REFLUX study: descriptive statistics and  cost-effectiveness  according to ITT and alternative methods for estimating the CACE. Follow-up period is  five years, and treatment switches are defined within the first year post randomisation.  Costs and INB numbers rounded to the nearest integer.}}
\label{TableReflux1}
\begin{tabular}{lrl}\hline \hline
          & Medical management & Laparoscopic surgery \\\hline
    N Assigned & $179$    & $178$ \\
    N (\%) Switched & $10$ $(8.3)$ & $67$  $(28.4)$ \\
N (\%) missing costs&   83  (46) & 83  (47)\\  
Mean (SD) observed cost in \pounds & $1258\  (1687)$ & $2971\  (1828)$ \\   
N (\%) missing QALYs& 91   (51) & 94     (53)\\  
 Mean (SD) observed QALYs  & $3.52\  (0.99)$ & $3.74 \ (0.90)$ \\

\hline 
\small{\it Baseline variables} &   & \\ \hline
 N (\%) missing  $\mbox{EQ5D}_{0} $& 6        (3)&  7  (4) \\
Mean (SD) observed $\mbox{EQ5D}_{0}$  & $0.72\  (0.25)$ & $0.71 \ (0.26)$ \\
 Correlation between costs and QALYs   & $-0.42$ & $-0.07$ \\
  Correlation  of costs and QALYs    & $-0.36$ & $-0.18$\\
by treatment received &&
\\ \hline \hline
        \multicolumn{3}{c}{\textbf{Incremental costs, QALYs and INB of surgery vs medicine}}  \\ \hline \hline
Outcome  &Method        & estimate  ( 95\% CI )\\
 \hline
Incremental cost &&\\
 \hline
& ITT& $1103\ \ (593, 1613)$\\
& 2sls &$1899\ (1073, 2724)$\\  
& 3sls &$1899\ (1073, 2724)$\\
&uBN &$  2960 \ (2026,   3998)$ \\
&uBGN &$ 2176   \ (1356,  3031)$\\
&BFL&$2030\  (1170,  2878)$\\
\hline
Incremental QALYs&&\\
 \hline
& ITT &  $0.295\ (0.002,  0.589)$ \\
 & 2sls& $0.516\ (0.103, 0.929)$\\    
 & 3sls& $0.516 \ (0.103, 0.929)$\\
& uBN &$0.568    \ (0.181, 0.971 )$ \\
 &uBGN &$ 0.268\ (-0.229,  0.759)$\\
&   BFL&  $ 0.511\ (0.121,  0.947)$\\
 \hline
INB& &\\
 \hline
& ITT &$7763\  (-1059, 16585)$\\
& 2sls& $13587\  ( 1101,26073)$ \\
& 3sls  & $13587\  (1002, 26173)$ \\
& uBN &$ 14091\  (2485, 26086)$\\ 
&uBGN &$5869\ (-9204, 20740)$ \\
& BFL &  $13340\  (1406,  26315)$\\
 \hline\hline
  \end{tabular}%
\end{table}
\end{center}

\begin{center}
\begin{table}
\caption{\footnotesize{CI Coverage rates (CR) and median width for incremental cost, QALYs, and INB, across scenarios with 30\% non-compliance, sample size $n= 100$ and moderate correlation between outcomes (with odd rows corresponding to positive correlation $\rho$ and even rows to negative). uBGN was not applied in settings with normal cost data.}}
\begin{footnotesize}
 \begin{tabular}{lrrrrrrrrrr} \hline
    \multicolumn{1}{c}{\textbf{}} & \multicolumn{2}{c}{\textbf{2sls}} & \multicolumn{2}{c}{\textbf{3sls}} & \multicolumn{2}{c}{\textbf{uBN}} &\multicolumn{2}{c}{\textbf{uBGN}} & \multicolumn{2}{c}{\textbf{BFL}} \\ \cline{2-11}
 \textbf{$Y_1\sim N$}  & \textbf{CR} & \textbf{CIW} & \textbf{CR} & \textbf{CIW} & \textbf{CR} & \textbf{CIW} & \textbf{CR} & \textbf{CIW}& \textbf{CR} & \textbf{CIW}\\ \hline
 \textbf{Cost} &  .952 & .228 &  .952 & .228 & .992 & .312 & &&.988 & .299 \\
 $\rho<0$   &  .952 & .229 &  .952 & .229 & .993 & .325 & &&.986 & .297 \\ \cline{2-11}

\textbf{QALYs}  &  .946 & .112 &  .946 & .112 & .988 & .155 & &&.950  & .121 \\
 $\rho<0$ & .950 & .113 & .950  & .113  & .992 & .163 & &&.950  & .121 \\ \cline{2-11}

\textbf{INB}  &  .988 & 405  &  .953 & 319   & .982 & 398   & &&.966 & 376 \\
  $\rho<0$  &.900   & 409  & .948 & 475   & .951 & 509   & &&.962 & 525 \\ \hline
  \textbf{$Y_1\sim G$} &       &       &       &       &       &       &       &       &  \\ \cline{1-11}
   
  \textbf{Cost}  & .952 & .756 & .952 & .756 & .955 & .815 &.941 & .818 & .954 & .823 \\
     $\rho<0$&  .942 & .759  & .942 & .759& .949 & .828 & .936 & .822  & .945 & .811 \\ \cline{2-11}

     \textbf{QALYs}   & .959 & .113  & .959 & .113 & .993 & .160& .960 & .122  & .960  & .122 \\
     $\rho<0$& .959 & .113 & .949 & .113 & .995 & .163& .954 & .122  & .954 & .122 \\ \cline{2-11}
    \textbf{INB}    & .982 & 829 & .948 & 696   & .958 & 764& .942 & 748   & .956 & 760 \\
    $\rho<0$& .914 & 833& .948 & 943   & .930  & 921 & .941 & 1019   & .951 & 1014 \\ \hline
    \textbf{$Y_1\sim IG$} &&       &       &       &       &       &       &       &       &  \\ \cline{1-11}

    \textbf{Cost} & .951 & .880& .951 & .880  & .958 & .949& .904 & .866 & .956 & .945 \\
    $\rho<0$  & .950  & .878 & .950  & .878   & .958 & .951& .905 & .864 & .954 & .932 \\ \cline{2-11}

     \textbf{QALYs}   & .945& .112 & .945 & .112 & .991 & .161& .944 & .120 & .999 & .206 \\
     $\rho<0$&.954 & .112 & .954 & .112& .993 & .161& .952 & .120 & .999 & .204 \\ \cline{2-11}

     \textbf{INB}  & .980  & 944 & .954 & 818   & .959 & 889&.917& 814  & .984 & 1001 \\
     $\rho<0$&  .917 & 942&.947 & 1049  & .934 & 1034  & .911 & 1041 & .971 & 1203 \\ \hline
\end{tabular}
\end{footnotesize}
 \label{CI30modcorr}%
\end{table}%
\end{center}

\begin{center}
\begin{table}
\caption{\small{RMSE for incremental Cost, QALYs and INB across scenarios with 30\% non-compliance, moderate correlation between outcomes and sample size $n=100$. uBGN was not applied in settings with normal cost data. Numbers for INB have been rounded to the nearest integer.}}
  \begin{tabular}{lcrrrrr} \hline
\centering

     \textbf{Cost distribution} & \textbf{} & \multicolumn{1}{c}{\textbf{$\rho$}}  & \multicolumn{1}{c}{\textbf{3sls (2sls)}} & \multicolumn{1}{c}{\textbf{uBN}}& \multicolumn{1}{c}{\textbf{uBGN}} & \multicolumn{1}{c}{\textbf{BFL}} \\ \hline
    \textbf{Normal} & \textbf{Cost} &       &       &       &       &  \\ \cline{2-7}
    \textbf{} & \textbf{} & 0.4   &  0.058 & 0.060 & & 0.059 \\
    \textbf{} & \textbf{} & -0.4  & 0.060& 0.062 & &0.061 \\ \cline{2-7}
    \textbf{} & \textbf{QALYs} &      &       &       &       &  \\ \cline{2-7}
    \textbf{} & \textbf{} & 0.4   &0.029 & 0.030 & &0.030 \\
    \textbf{} & \textbf{} & -0.4  & 0.029 & 0.030 & &0.030\\ \cline{2-7}
    \textbf{} & \textbf{INB} &       &       &       &       &  \\ \cline{2-7}
    \textbf{} & \textbf{} & 0.4   & 83  & 84  & &87 \\
    \textbf{} & \textbf{} & -0.4  & 125 & 127 & &125 \\ \hline
    \textbf{Gamma} & \textbf{Cost} &       &       &       &       &  \\ \cline{2-7}
    \textbf{} & \textbf{} & 0.4    & 0.198  & 0.202 & 0.212& 0.202 \\
    \textbf{} & \textbf{} & -0.4  &  0.200 & 0.204 & 0.212& 0.203 \\ \cline{2-7}
    \textbf{} & \textbf{QALYs} &       &       &       &       &  \\ \cline{2-7}
    \textbf{} & \textbf{} & 0.4   & 0.030 & 0.030 &0.030& 0.029 \\
    \textbf{} & \textbf{} & -0.4  &  0.029   & 0.030 & 0.030& 0.030 \\ \cline{2-7}
    \textbf{} & \textbf{INB} &       &       &       &       &  \\ \cline{2-7}
    \textbf{} & \textbf{} & 0.4   &  181   & 184 &193  & 184 \\
    \textbf{} & \textbf{} & -0.4  & 246   & 251 &261  & 252 \\ \hline
    \textbf{Inverse} & \textbf{Cost} &       &       &       &       &  \\ \cline{2-7}
    \textbf{Gaussian} & \textbf{} & 0.4   & 0.230 & 0.232 &0.252& 0.232 \\
    \textbf{} & \textbf{} & -0.4  & 0.230 & 0.232 &0.250& 0.232 \\ \cline{2-7}
    \textbf{} & \textbf{QALYs} &       &       &       &       &  \\ \cline{2-7}
    \textbf{} & \textbf{} & 0.4   & 0.029& 0.030 &0.030& 0.030\\
    \textbf{} & \textbf{} & -0.4  & 0.029 & 0.030 &0.030& 0.030 \\ \cline{2-7}
    \textbf{} & \textbf{INB} &       &       &       &       &  \\ \cline{2-7}
    \textbf{} & \textbf{} & 0.4   & 211   & 214   & 231&214 \\
    \textbf{} & \textbf{} & -0.4  & 273   & 278 &296  & 278 \\ \hline
    \end{tabular}%
  \label{rmse30N100}%
\end{table}%
\end{center}

\clearpage
\appendix
\setcounter{table}{0}
\renewcommand{\thetable}{A\arabic{table}}
\section{ Sensitivity analyses for the REFLUX RCT}

Our motivating example trial  had over 50\% missing cost-effectiveness outcomes, with a few missing data in the baseline covariate.  As primary analysis, we performed MI and full Bayesian analyses, that intrinsically obtain a predictive posterior distribution for these missing values, and reported the results of these analyses in the main text.

As a sensitivity to the MAR assumption, we present here the complete case analysis  (N=172)  of the REFLUX RCT,  adjusting for baseline EQ5D. The results are valid if conditional on baseline EQ5D, the missing data mechanism is independent of the outcomes.
\begin{center}
\begin{table}[h]
\caption{\small{Estimated cost-effectiveness  for the REFLUX RCT, over a time horizon of five years according to ITT and alternative methods for estimating the CACE based on complete cases. Treatment switches are defined within the first year post randomisation.
 Costs and INB numbers rounded to the nearest integer.}}
\label{TableRefluxCCA}
\centering
\begin{tabular}{rrr}\hline \hline
        \multicolumn{3}{c}{\textbf{Incremental costs, QALYs and INB of surgery vs medicine}}  \\ \hline \hline
Outcome  &Method        & estimate  ( 95\% CI )\\
 \hline
Incremental cost &&\\
 \hline
& ITT& $1626\  (1098, 2153)$\\
& 2sls &$2102\ (1484, 2720)$\\
& 3sls &$2102\ (1484, 2720)$\\
&uBN &$2112\  (1401,  2879)$ \\
&uBGN &$ 2195  \ (1481,  3065)$\\
&BFL&$2081\ (1457, 2705)$\\

\hline
Incremental QALYs&&\\
 \hline
& ITT &  $0.319\  (-0.055,  0.694)$ \\
& 2sls& $0.412\ (0.137, 0.688)$\\
& 3sls& $0.412\ (0.137, 0.688)$\\
& uBN &$0.413\ (0.136,   0.704)$ \\
&uBGN &$ 0.285\ (-0.066,  0.629)$\\
& BFL&  $0.414   \ (0.135 ,  0.703)$
\\
 \hline
INB& &\\
 \hline
& ITT & $ 7948.\ (-3259, 19155)$ \\
& 2sls& $10275\ (1997, 18554)$ \\
& 3sls  & $10275\  (1828, 18724)$ \\
& uBN &$10280, \ (1947, 18980)$ \\
&uBGN &$ 6362\ -4497, 16941)$\\
& BFL & $ 10353\ (1789, 19165)$ \\
 \hline
  \end{tabular}%
\end{table}
\end{center}   

As can be seen in Table \ref{TableRefluxCCA}, the conclusions each model reaches are not substantially changed from those under MAR (either using MI or full Bayesian), presented in the main text. 

\subsection{Sensitivity analysis of the BFL to choice of priors}

We now study the sensitivity of the  BFL model to the choice of priors for the parameters in the model. 
Recall that this model can be written as 
 \begin{equation}
\begin{array}{l}
\left(\begin{array}{c}
D_{i}\\ Y_{1i} \\ Y_{2i}\end{array}\right)
\sim N\left\{\left(\begin{array}{c}
\mu_{0i}\\
\mu_{1i}\\
\mu_{2i}\end{array}\right),  \Sigma\right\}
\end{array}\ \ \ \ \ \ \ \
\begin{array}{l}
\Sigma= \left(\begin{array}{ccc} \s_{0} & s_{01} & s_{02} \\  s_{01}& \s_{1} & s_{12} \\
 s_{02}& s_{12} & \s_{2}
 \end{array} \right)
\end{array}
 \end{equation}
where $s_{ij}=\mbox{cov}(Y_i,Y_j)$, and
\begin{eqnarray}\label{MVN_IVBayes}
\nonumber\mu_{0i} &=& \beta_{0,0} + \beta_{1,0}Z_i \\
\nonumber\mu_{1i} &=& \beta_{0,1}+ \beta_{1,1} \beta_{1,0}Z_i  \\
\mu_{2i}&=&  \beta_{0,2}+\beta_{1,2}\beta_{1,0}Z_i
\end{eqnarray}

Originally, we chose normal priors for the regression coefficients, $\beta_{m,j}\sim N(0, 10^2)$, for $j \in\{0,1,2\}, m\in\{0,1\}$, and a Wishart prior for the inverse of $\Sigma$ \citep{GelmanHill}.

Here, we first vary the prior distribution for $\Sigma$, and assume a structured covariance matrix (see \cite{Congdon2007} example 5.8, and \cite{BUGSbook}, example 9.1.4).   Thus, we write:
 \begin{equation}\label{structureprec}
\begin{array}{l}
\Sigma= \left(\begin{array}{ccc} s_{00}s_{00} & \rho_1 s_{00}s_{11} &\rho_2 s_{00}s_{22}\\
 \rho_1 s_{00}s_{11} & s_{11}s_{11} &\rho_3 s_{11}s_{22} \\
\rho_2 s_{00}s_{22}& \rho_3 s_{11}s_{22} & s_{22}s_{22}
 \end{array} \right)
\end{array}
\end{equation}
and  assumed   the following priors
\begin{equation}\label{MVN_IVBayes}
s_{ij}\sim N(0, 10^2), \qquad \rho_j \sim \mbox{Unif}[-1,1]
\end{equation}
In a secondary sensitivity analysis, we assume a Wishart prior for $\Sigma$ as before, but used uniform priors for the regression coefficients, $\beta_{m,j}\sim \mbox{Unif}[-10,10]$, for $j \in\{0,1,2\}, m\in\{0,1\}$.  
We did these changes on both available cases and complete cases.
The results corresponding to INB are reported in Table \ref{tab:refluxinbsensitivity}. 

\begin{table}[h]
\caption{Results from sensitivity analysis of the BFL  to prior specifications for the INB of the motivating example: CEA within the REFLUX trial. INB reported to the nearest integer, in Great British Pounds (GBP).}
\begin{center}
\begin{threeparttable}
\begin{tabular}{rrrr}\hline
Data & Priors &INB estimate  & 95\% CI \\
    \hline
Available cases& Structured precision matrix\tnote{1} &  $10421$& $(1878, 19187)$\\ 
&normal priors for  $\beta_{m,j}$ & &\\
Available cases & Wishart prior for the precision  &$10405$ & $(1733,19330)$\\
&Unif prior for  $\beta_{m,j}$ & & \\
Complete cases& Structured precision matrix &  10425 &$(1792, 19156)$ \\ 
&normal priors for  $\beta_{m,j}$ &   &\\
Complete cases & Wishart prior for the precision  &  $10428$ &$(1894,  19279)$ \\
&Unif prior for  $\beta_{m,j}$ & & \\
   \hline
\end{tabular}%
\begin{tablenotes}
\item[1] Following the specifications of equation \eqref{MVN_IVBayes} in this Appendix.
\end{tablenotes}
\end{threeparttable}
\end{center}
\label{tab:refluxinbsensitivity}
\end{table}

\section{Supplementary Results from the simulations}
\begin{figure}
\begin{center}
\caption{\footnotesize{Median Bias for scenarios with  70\% non-compliance and sample sizes of (a) $n=100$ (top) and (b)  $n=1000$. Results are stratified by cost distribution, and correlation between cost and QALYs. The dotted line represents zero bias. Results for 2sls (not plotted) are identical to those for 3sls; uBGN was not applied to Normal cost data.}}\label{70Biasplot}
\vskip -.3ex
\subfloat[n=100]{\includegraphics[scale=.45]{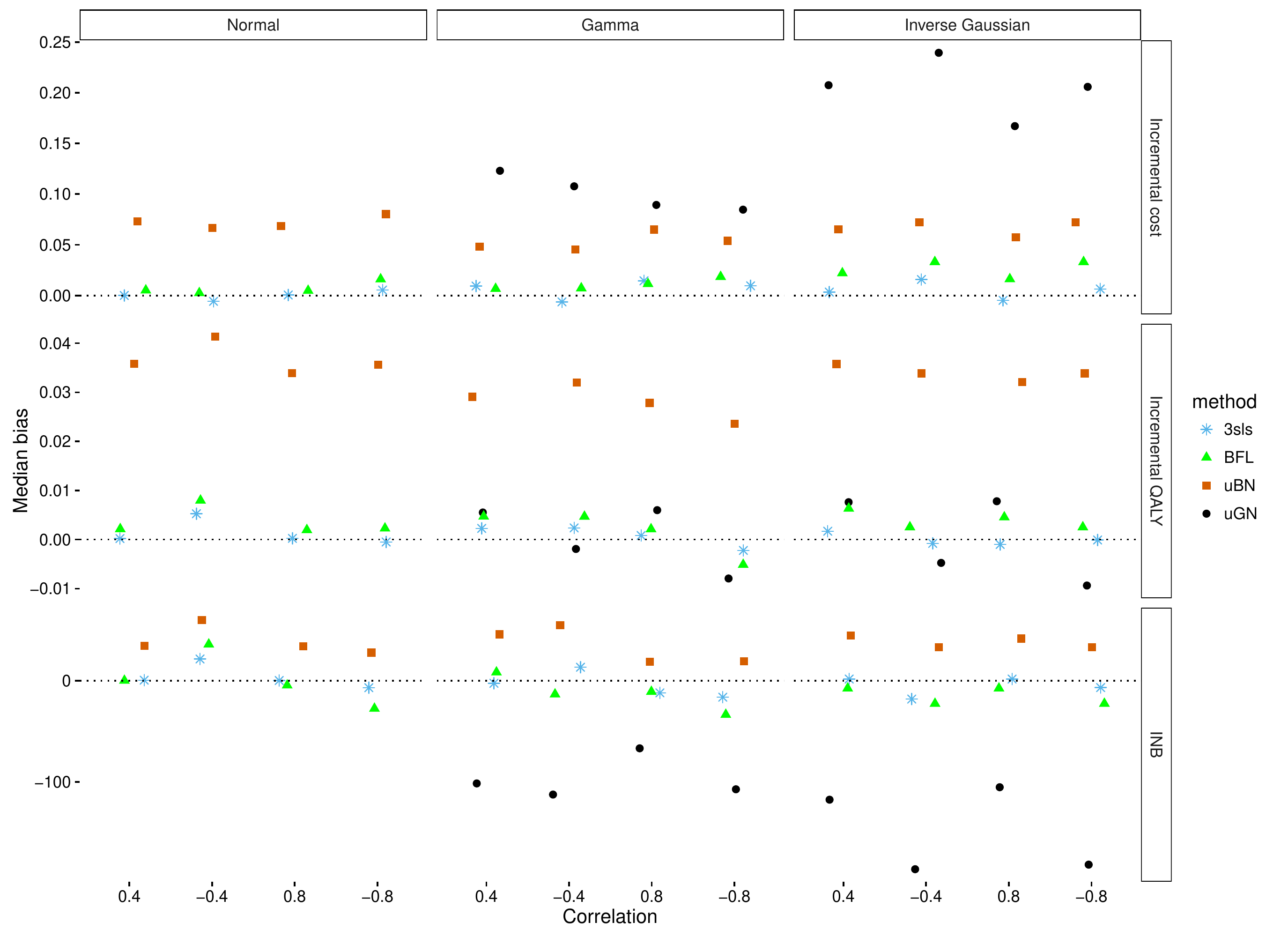}}\\ \vskip -.5ex
\subfloat[n=1000]{\includegraphics[scale=.45]{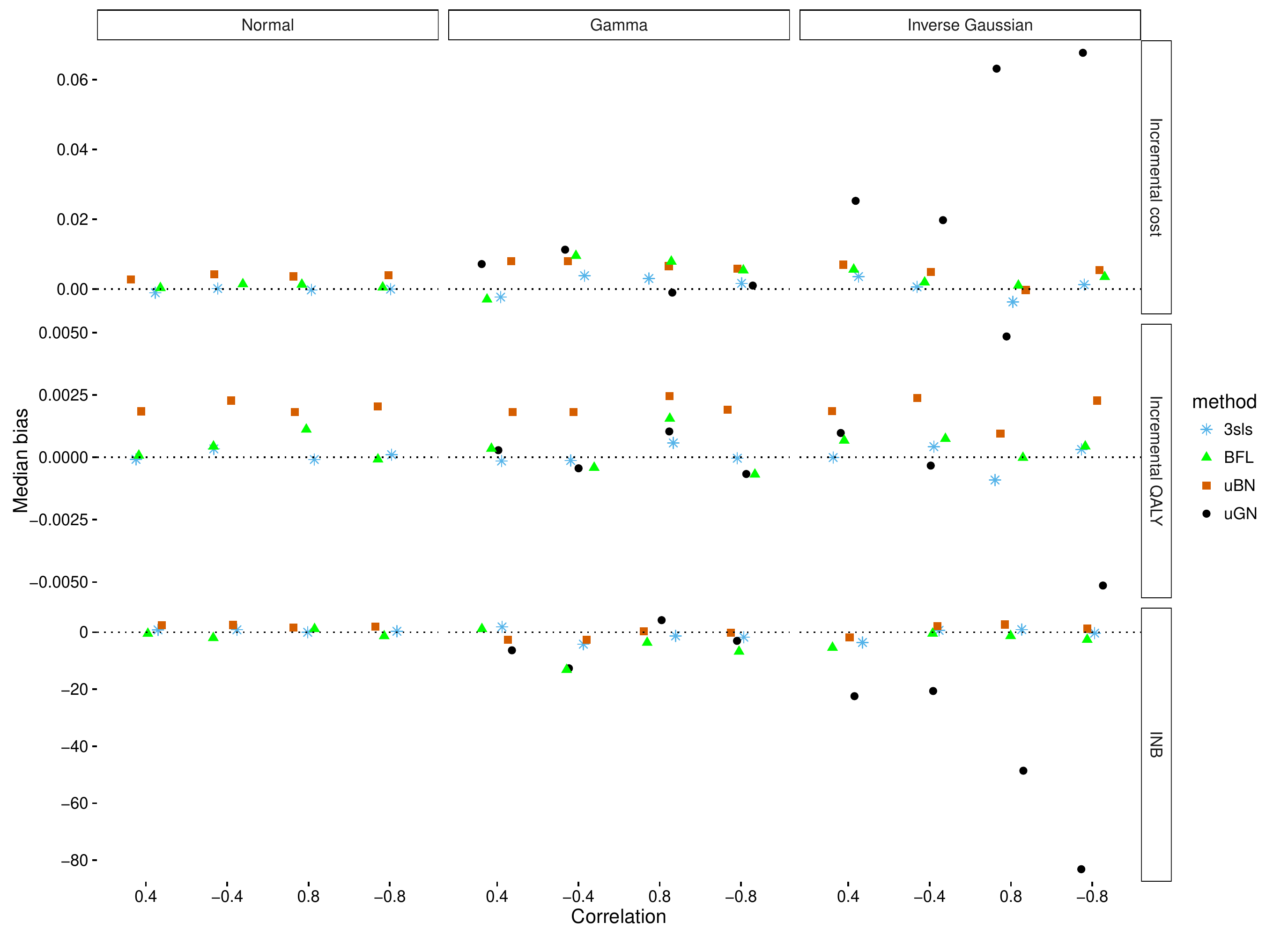}}
\vskip -.7ex
\end{center}
\end{figure}

\begin{sidewaystable}
  \centering
  \caption{CI Coverage rates and median width for incremental Cost, QALYs, and INB, across scenarios with 30\% non-compliance, moderate correlation between outcomes and sample size $n= 1000$. uBGN was not applied in settings with normal cost data.}
\begin{tabular}{rcrrrrrrrrrrr} \hline
    \multicolumn{1}{l}{\textbf{}} & \textbf{} & \multicolumn{1}{c}{\textbf{}}  & \multicolumn{2}{c}{\textbf{2sls}} &\multicolumn{2}{c}{\textbf{3sls}} & \multicolumn{2}{c}{\textbf{uBN}}& \multicolumn{2}{c}{\textbf{uBGN}} & \multicolumn{2}{c}{\textbf{BFL}} \\\cline{4-13}
  \textbf{Cost distribution} & \textbf{} & \textbf{$\rho$}   & \textbf{coverage} & \textbf{width} & \textbf{coverage} & \textbf{width} & \textbf{coverage} & \textbf{width} & \textbf{coverage} & \textbf{width} & \textbf{coverage} & \textbf{width} \\ \hline
 \multicolumn{1}{l}{\textbf{Normal}} & \textbf{Cost} &   &    &       &       &       &       &       &       &       &  \\ \cline{2-13}
    \multicolumn{1}{l}{\textbf{}} & \textbf{} & 0.4 &    0.949 & 0.072 &   0.949 & 0.072 & 0.989 & 0.094 & && 0.956 & 0.075 \\
    \multicolumn{1}{l}{\textbf{}} & \textbf{} & -0.4  & 0.957 & 0.072 & 0.957 & 0.072 & 0.992 & 0.097 & &&0.963 & 0.075 \\ \cline{2-13}
    \multicolumn{1}{l}{\textbf{}} & \textbf{QALYs} &     &  &       &       &       &       &       &       &       &  \\ \cline{2-13}
    \multicolumn{1}{l}{\textbf{}} & \textbf{} & 0.4  &0.942 & 0.036  & 0.942 & 0.036 & 0.982 & 0.046 & &&0.937 & 0.035 \\
    \multicolumn{1}{l}{\textbf{}} & \textbf{} & -0.4  &  0.947 & 0.036 & 0.947 & 0.036 & 0.991 & 0.048 & &&0.944 & 0.035 \\
\hline
    \end{tabular}%
  \label{CI30modcorr1000}%
\end{sidewaystable}%

\clearpage
\begin{sidewaystable}[h]
  \centering
  \caption{CI Coverage rates and median width for incremental Cost, QALYs, and INB, across scenarios with 70\% non-compliance, moderate correlation between outcomes and sample size $n=100$. uBGN was not applied in settings with normal cost data.}
    \begin{tabular}{rcrrrrrrrrrrr} \hline
    \multicolumn{1}{l}{\textbf{}} & \textbf{} & \multicolumn{1}{c}{\textbf{}}& \multicolumn{2}{c}{\textbf{2sls}}  & \multicolumn{2}{c}{\textbf{3sls}} & \multicolumn{2}{c}{\textbf{uBN}}& \multicolumn{2}{c}{\textbf{uBGN}} & \multicolumn{2}{c}{\textbf{BFL}} \\ \cline{4-13}
 \textbf{Cost distribution} & \textbf{} & \textbf{$\rho$}   & \textbf{coverage} & \textbf{width} & \textbf{coverage} & \textbf{width} & \textbf{coverage} & \textbf{width} & \textbf{coverage} & \textbf{width} & \textbf{coverage} & \textbf{width} \\ \hline
     \multicolumn{1}{l}{\textbf{Normal}} & \textbf{Cost} &       &       &       &       &       &       &       &       &  \\ \cline{2-13}
\multicolumn{1}{l}{\textbf{}} & \textbf{} & 0.4   & 0.968 & 0.550 & 0.968 & 0.550& 0.988 & 1.33 && & 0.991 & 1.13 \\
    \multicolumn{1}{l}{\textbf{}} & \textbf{} & -0.4  &  0.966 & 0.545& 0.966 & 0.545& 0.990  & 1.30   & &&0.985 & 1.08 \\ \cline{2-13}
    \multicolumn{1}{l}{\textbf{}} & \textbf{QALYs} & &       &       &       &       &       &       &       &       &  \\ \cline{2-13}
    \multicolumn{1}{l}{\textbf{}} & \textbf{} & 0.4     & 0.968 & 0.271 & 0.968 & 0.271   & 0.988 & 0.665 & &&0.964 & 0.445 \\
    \multicolumn{1}{l}{\textbf{}} & \textbf{} & -0.4   & 0.964 & 0.269 & 0.964 & 0.269 & 0.986 & 0.665 & &&0.954 & 0.435 \\
\hline
    \end{tabular}\label{CI70modcorr}
\end{sidewaystable}%

\begin{sidewaystable}
  \centering
  \caption{CI Coverage rates and median width for incremental Cost, QALYs, and INB, across scenarios with 70\% non-compliance, moderate correlation between outcomes and sample size $n=1000$. uBGN was not applied in settings with normal cost data.}
    \begin{tabular}{rcrrrrrrrrrrr} \hline
    \multicolumn{1}{l}{\textbf{}} & \textbf{} & \multicolumn{1}{c}{\textbf{}} & \multicolumn{2}{c}{\textbf{2sls}} &\multicolumn{2}{c}{\textbf{3sls}} & \multicolumn{2}{c}{\textbf{uBN}}& \multicolumn{2}{c}{\textbf{uBGN}}  & \multicolumn{2}{c}{\textbf{BFL}} \\ \cline{4-13}
 \textbf{Cost distribution} & \textbf{} & \textbf{$\rho$}   & \textbf{coverage} & \textbf{width} & \textbf{coverage} & \textbf{width} & \textbf{coverage} & \textbf{width} & \textbf{coverage} & \textbf{width} & \textbf{coverage} & \textbf{width} \\ \hline
     \multicolumn{1}{l}{\textbf{Normal}} & \textbf{Cost} &       &       &       &       &       &       &  &&     &       &  \\ \cline{2-13}
    \multicolumn{1}{l}{\textbf{}} & \textbf{} & 0.4  &0.954 & 0.169& 0.954 & 0.169 & 0.986 & 0.223 & && 0.958 & 0.178 \\
    \multicolumn{1}{l}{\textbf{}} & \textbf{} & -0.4  & 0.953 & 0.169& 0.953 & 0.169& 0.991 & 0.230  & &&0.960  & 0.179 \\ \cline{2-13}
    \multicolumn{1}{l}{\textbf{}} & \textbf{QALYs} & &&      &       &       &       &       &       &       &       &  \\ \cline{2-13}
    \multicolumn{1}{l}{\textbf{}} & \textbf{} & 0.4    & 0.946 & 0.083 & 0.946 & 0.083  & 0.986 & 0.111 &&& 0.942 & 0.084\\
    \multicolumn{1}{l}{\textbf{}} & \textbf{} & -0.4   & 0.953 & 0.083   & 0.953 & 0.083  & 0.992 & 0.115 &&& 0.947 & 0.083\\
\hline
    \end{tabular}%
  \label{CI70modcorr1000}%
\end{sidewaystable}%

\begin{sidewaystable}[h]
  \centering
  \caption{CI Coverage rates and median width for incremental Cost, QALYs, and INB, across scenarios with 30\% non-compliance, high correlation between outcomes and sample size $n=100$. uBGN was not applied in settings with normal cost data.}
    \begin{tabular}{lcrrrrrrrrrrrrr} \hline
    \textbf{} & \textbf{} & \multicolumn{1}{c}{\textbf{}} &  \multicolumn{2}{c}{\textbf{2sls}} &  \multicolumn{2}{c}{\textbf{3sls}} & \multicolumn{2}{c}{\textbf{uBN}} &\multicolumn{2}{c}{\textbf{uBGN}} & \multicolumn{2}{c}{\textbf{BFL}} \\ \cline{4-13}
    \textbf{Cost distribution} & \textbf{} & \textbf{$\rho$}   & \textbf{coverage} & \textbf{width} & \textbf{coverage} & \textbf{width} & \textbf{coverage} & \textbf{width} & \textbf{coverage} & \textbf{width}& \textbf{coverage} & \textbf{width} \\ \hline
    \textbf{Normal} & \textbf{Cost} &    &&   &       &       &       &       &       &       &       &  \\ \cline{2-13}
    \textbf{} & \textbf{} & 0.8   & 0.951 & 0.229&  0.951 & 0.229 & 0.984 & 0.299 &&& 0.987 & 0.299 \\
    \textbf{} & \textbf{} & -0.8  &  0.946 & 0.228 &  0.946 & 0.228& 0.992 & 0.324 & &&0.986 & 0.295 \\ \cline{2-13}
    \textbf{} & \textbf{QALYs} &&&       &       &       &       &       &       &       &       &  \\ \cline{2-13}
    \textbf{} & \textbf{} & 0.8   & 0.954 & 0.113  & 0.954 & 0.113  & 0.985 & 0.148 & &&0.958 & 0.122 \\
    \textbf{} & \textbf{} & -0.8  &  0.949 & 0.112  &  0.949 & 0.112  & 0.992 & 0.162 & &&0.952 & 0.121 \\
\hline
    \end{tabular}%
  \label{CI30highcorr}%
\end{sidewaystable}%

\begin{sidewaystable}[h]
  \centering
  \caption{CI Coverage rates and median width for incremental Cost, QALYs, and INB, across scenarios with 30\% non-compliance, high correlation between outcomes and sample size $n=1000$.uBGN was not applied in settings with normal cost data.}
    \begin{tabular}{rcrrrrrrrrrrr} \hline
    \multicolumn{1}{l}{\textbf{}} & \textbf{} & \multicolumn{1}{c}{\textbf{}} &  \multicolumn{2}{c}{\textbf{2sls}}&  \multicolumn{2}{c}{\textbf{3sls}} & \multicolumn{2}{c}{\textbf{uBN}}&\multicolumn{2}{c}{\textbf{uBGN}}  & \multicolumn{2}{c}{\textbf{BFL}} \\ \cline{4-13}
    \multicolumn{1}{c}{\textbf{Cost distribution}} & \textbf{} & \multicolumn{1}{c}{\textbf{$\rho$}} & \multicolumn{1}{c}{\textbf{coverage}} & \multicolumn{1}{c}{\textbf{width}} & \multicolumn{1}{c}{\textbf{coverage}} & \multicolumn{1}{c}{\textbf{width}} & \multicolumn{1}{c}{\textbf{coverage}} & \multicolumn{1}{c}{\textbf{width}} & \multicolumn{1}{c}{\textbf{coverage}} & \multicolumn{1}{c}{\textbf{width}} & \textbf{coverage} & \textbf{width}\\ \hline
    \multicolumn{1}{l}{\textbf{Normal}} & \textbf{Cost} &       &       &       &       &       &       &       &       &  \\ \cline{2-13}
    \multicolumn{1}{l}{\textbf{}} & \textbf{} & 0.8   &  .949 & 0.072 &  0.949 & 0.072 & 0.985 & 0.089 &&& 0.951 & 0.074 \\
    \multicolumn{1}{l}{\textbf{}} & \textbf{} & -0.8  & 0.954 & 0.072 &  0.954 & 0.072& 0.996 & 0.097 & &&0.955 & 0.074 \\ \cline{2-13}
    \multicolumn{1}{l}{\textbf{}} & \textbf{QALYs} &  &&     &       &       &       &       &       &       &       &  \\ \cline{2-13}
    \multicolumn{1}{l}{\textbf{}} & \textbf{} & 0.8   & 0.943 & 0.036&  0.943 & 0.036 0 & 0.988 & 0.044 & &&0.938 & 0.035 \\
    \multicolumn{1}{l}{\textbf{}} & \textbf{} & -0.8  &  0.952 & 0.036&  0.952 & 0.036   & 0.992 & 0.048 & &&0.951 & 0.036 \\
\hline
    \end{tabular}%
  \label{CI30highcorr1000}%
\end{sidewaystable}%

\begin{sidewaystable}[h]
  \centering
  \caption{CI Coverage rates and median width for incremental Cost, QALYs, and INB, across scenarios with 70\% non-compliance, high correlation between outcomes and sample size $n=100$. uBGN was not applied in settings with normal cost data.}
    \begin{tabular}{rcrrrrrrrrrrr} \hline
    \multicolumn{1}{l}{\textbf{}} & \textbf{} & \multicolumn{1}{c}{\textbf{}}& \multicolumn{2}{c}{\textbf{2sls}}& \multicolumn{2}{c}{\textbf{3sls}} & \multicolumn{2}{c}{\textbf{uBN}}  & \multicolumn{2}{c}{\textbf{uBGN}} & \multicolumn{2}{c}{\textbf{BFL}} \\ \cline{4-13}
    \multicolumn{1}{c}{\textbf{Cost distribution}} & \textbf{} & \multicolumn{1}{c}{\textbf{$\rho$}} & \multicolumn{1}{c}{\textbf{coverage}} & \multicolumn{1}{c}{\textbf{width}} & \multicolumn{1}{c}{\textbf{coverage}} & \multicolumn{1}{c}{\textbf{width}} & \multicolumn{1}{c}{\textbf{coverage}} & \multicolumn{1}{c}{\textbf{width}} & \multicolumn{1}{c}{\textbf{coverage}} & \multicolumn{1}{c}{\textbf{width}} & \textbf{coverage} & \textbf{width}\\ \hline
    \multicolumn{1}{l}{\textbf{Normal}} & \textbf{Cost} &    &&   &       &       &       &       &       &       &       &  \\ \cline{2-13}
    \multicolumn{1}{l}{\textbf{}} & \textbf{} & 0.8   &  0.972 & 0.546& 0.972 & 0.546& 0.982 & 1.29  &&& 0.988 & 1.09 \\
    \multicolumn{1}{l}{\textbf{}} & \textbf{} & -0.8  &0.970  & 0.550  &0.970  & 0.550  & 0.99  & 1.36  & &&0.988 & 1.07 \\ \cline{2-13}
    \multicolumn{1}{l}{\textbf{}} & \textbf{QALYs} &&&       &       &       &       &       &       &       &       &  \\ \cline{2-13}
    \multicolumn{1}{l}{\textbf{}} & \textbf{} & 0.8   & 0.97  & 0.269& 0.97  & 0.269 & 0.982 & 0.64  & &&0.966 & 0.443 \\
    \multicolumn{1}{l}{\textbf{}} & \textbf{} & -0.8 & 0.966 & 0.272& 0.966 & 0.272  & 0.989 & 0.662&& & 0.960  & 0.446 \\
\hline
    \end{tabular}%
  \label{CI70highcorr}%
\end{sidewaystable}%

\begin{sidewaystable}[h]
  \centering
  \caption{CI Coverage rates and median width for incremental Cost, QALYs, and INB, across scenarios with 70\% non-compliance, high correlation between outcomes and sample size $n=1000$. uBGN was not applied in settings with normal cost data.}
    \begin{tabular}{rcrrrrrrrrrrr} \hline
    \multicolumn{1}{l}{\textbf{}} & \textbf{} & \multicolumn{1}{c}{\textbf{}}  & \multicolumn{2}{c}{\textbf{2sls}}& \multicolumn{2}{c}{\textbf{3sls}} & \multicolumn{2}{c}{\textbf{uBN}}& \multicolumn{2}{c}{\textbf{uBGN}} & \multicolumn{2}{c}{\textbf{BFL}} \\ \cline{4-13}
    \multicolumn{1}{c}{\textbf{Cost distribution}} & \textbf{} & \multicolumn{1}{c}{\textbf{$\rho$}} & \multicolumn{1}{c}{\textbf{coverage}} & \multicolumn{1}{c}{\textbf{width}} & \multicolumn{1}{c}{\textbf{coverage}} & \multicolumn{1}{c}{\textbf{width}} & \multicolumn{1}{c}{\textbf{coverage}} & \multicolumn{1}{c}{\textbf{width}} & \multicolumn{1}{c}{\textbf{coverage}} & \multicolumn{1}{c}{\textbf{width}} & \textbf{coverage} & \textbf{width}\\ \hline
    \multicolumn{1}{l}{\textbf{Normal}} & \textbf{Cost} &  &&     &       &       &       &       &       &       &       &  \\ \cline{2-13}
    \multicolumn{1}{l}{\textbf{}} & \textbf{} & 0.8    & 0.959 & 0.169 & 0.959 & 0.169 & 0.983 & 0.211 & &&0.959 & 0.175 \\
    \multicolumn{1}{l}{\textbf{}} & \textbf{} & -0.8   &0.952 & 0.169 & 0.952 & 0.169 & 0.991 & 0.230  &  &&0.954 & 0.177 \\ \cline{2-13}
    \multicolumn{1}{l}{\textbf{}} & \textbf{QALYs} &       &       &      && &       &       &       &       &       &  \\ \cline{2-13}
    \multicolumn{1}{l}{\textbf{}} & \textbf{} & 0.8   & 0.957 & 0.083  & 0.957 & 0.083 & 0.983 & 0.106 &  &&0.951 & 0.084 \\
    \multicolumn{1}{l}{\textbf{}} & \textbf{} & -0.8   &  0.954 & 0.083 & 0.954 & 0.083 & 0.991 & 0.115 & && 0.951 & 0.085 \\
\hline
    \end{tabular}%
  \label{CI70highcorr1000}%
\end{sidewaystable}%

\begin{table}[h]
  \centering
  \caption{RMSE for incremental Cost, QALYs and INB across scenarios with 30\% non-compliance, high correlation between outcomes and sample size $n=100$. uBGN was not applied in settings with normal cost data. Numbers for INB have been rounded to the nearest integer.}
    \begin{tabular}{lcrrrrr} \hline
    \textbf{Cost distribution} & \textbf{} & \multicolumn{1}{c}{\textbf{$\rho$}}  & \multicolumn{1}{c}{\textbf{3sls  (2sls)}} & \multicolumn{1}{c}{\textbf{uBN}}& \multicolumn{1}{c}{\textbf{uBGN}} & \multicolumn{1}{c}{\textbf{BFL}} \\ \hline
    \textbf{Normal} & \textbf{Cost} &       &       &       &       &  \\ \cline{2-7}
    \textbf{} & \textbf{} & 0.8   &  0.059 & 0.060 && 0.060\\
    \textbf{} & \textbf{} & -0.8 & 0.059 & 0.061 & &0.060\\ \cline{2-7}
    \textbf{} & \textbf{QALYs} &       &       &       &       &  \\ \cline{2-7}
    \textbf{} & \textbf{} & 0.8   & 0.029 & 0.030 & &0.030 \\
    \textbf{} & \textbf{} & -0.8  & 0.029 & 0.030 & &0.030 \\ \cline{2-7}
    \textbf{} & \textbf{INB} &       &       &       &       &  \\
\hline
    \end{tabular}%
  \label{rmse30N100_high}%
\end{table}%

\begin{table}[h]
  \centering
  \caption{RMSE for incremental Cost, QALYs and INB across scenarios with 30\% non-compliance, moderate correlation between outcomes and sample size $n=1000$. uBGN was not applied in settings with normal cost data. Numbers for INB have been rounded to the nearest integer.}
    \begin{tabular}{lcrrrrr} \hline
    \textbf{Cost distribution} & \textbf{} & \multicolumn{1}{c}{\textbf{$\rho$}}  & \multicolumn{1}{c}{\textbf{3sls (2sls)}} & \multicolumn{1}{c}{\textbf{uBN}}& \multicolumn{1}{c}{\textbf{uBGN}} & \multicolumn{1}{c}{\textbf{BFL}} \\ \hline
    \textbf{Normal} & \textbf{Cost} &       &       &       &       &  \\ \cline{2-7}
    \textbf{} & \textbf{} & 0.4   &  0.018 & 0.018 && 0.018 \\
    \textbf{} & \textbf{} & -0.4  &  0.018 & 0.018 & &0.018 \\ \cline{2-7}
    \textbf{} & \textbf{QALYs} &       &       &       &       &  \\ \cline{2-7}
    \textbf{} & \textbf{} & 0.4   &  0.009& 0.009 & &0.009 \\
    \textbf{} & \textbf{} & -0.4  & 0.009& 0.009 & &0.009 \\ \cline{2-7}
    \textbf{} & \textbf{INB} &       &       &       &       &  \\
\hline
    \end{tabular}%
  \label{rmse30N1000}%
\end{table}%

\begin{table}[h]
  \centering
  \caption{RMSE for incremental Cost, QALYs and INB across scenarios with 30\% non-compliance, high correlation between outcomes and sample size $n=1000$. uBGN was not applied in settings with normal cost data. Numbers for INB have been rounded to the nearest integer.}
    \begin{tabular}{lcrrrrr} \hline
    \textbf{Cost distribution} & \textbf{} & \multicolumn{1}{c}{\textbf{$\rho$}}  & \multicolumn{1}{c}{\textbf{3sls (2sls)}} & \multicolumn{1}{c}{\textbf{uBN}} & \multicolumn{1}{c}{\textbf{uBGN}}& \multicolumn{1}{c}{\textbf{BFL}} \\ \hline
    \textbf{Normal} & \textbf{Cost} &       &       &       &       &  \\ \cline{2-7}
    \textbf{} & \textbf{} & 0.8    & 0.019 & 0.019 &&0.019 \\
    \textbf{} & \textbf{} & -0.8   & 0.018 & 0.018 & &0.018 \\ \cline{2-7}
    \textbf{} & \textbf{QALYs} &       &       &       &       &  \\ \cline{2-7}
    \textbf{} & \textbf{} & 0.8    & 0.009 & 0.009 & &0.009 \\
    \textbf{} & \textbf{} & -0.8   & 0.009 & 0.009 & &0.009 \\ \cline{2-7}
    \textbf{} & \textbf{INB} &       &       &       &       &  \\
\hline
    \end{tabular}%
  \label{rmse30N1000_high}%
\end{table}%

\begin{table}[h]
  \centering
  \caption{RMSE for incremental Cost, QALYs and INB across scenarios with 70\% non-compliance, moderate correlation between outcomes and sample size $n=100$. uBGN was not applied in settings with normal cost data. Numbers for INB have been rounded to the nearest integer.}
    \begin{tabular}{lcrrrrr} \hline
    \textbf{Cost distribution} & \textbf{} & \multicolumn{1}{c}{\textbf{$\rho$}}  & \multicolumn{1}{c}{\textbf{3sls (2sls)}} & \multicolumn{1}{c}{\textbf{uBN}} & \multicolumn{1}{c}{\textbf{uBGN}}& \multicolumn{1}{c}{\textbf{BFL}} \\ \hline
    \textbf{Normal} & \textbf{Cost} &       &       &       &       &  \\ \cline{2-7}
    \textbf{} & \textbf{} & 0.4   &  0.145  & 0.207 && 0.181 \\
    \textbf{} & \textbf{} & -0.4  &   0.146 & 0.203 && 0.182 \\ \cline{2-7}
    \textbf{} & \textbf{QALYs} &       &       &       &       &  \\ \cline{2-7}
    \textbf{} & \textbf{} & 0.4   &  0.074 & 0.104 & &0.089 \\
    \textbf{} & \textbf{} & -0.4  &  0.074 & 0.109 & &0.088 \\ \cline{2-7}
    \textbf{} & \textbf{INB} &       &       &       &       &  \\
\hline
    \end{tabular}%
  \label{rmse70N100}%
\end{table}%

\begin{table}[h]
  \centering
  \caption{RMSE for incremental Cost, QALYs and INB across scenarios with 70\% non-compliance, high correlation between outcomes and sample size $n=100$. uBGN was not applied in settings with normal cost data. Numbers for INB have been rounded to the nearest integer.}
    \begin{tabular}{lcrrrrr} \hline
    \textbf{Cost distribution} & \textbf{} & \multicolumn{1}{c}{\textbf{$\rho$}}  & \multicolumn{1}{c}{\textbf{3sls (2sls)}} & \multicolumn{1}{c}{\textbf{uBN}} & \multicolumn{1}{c}{\textbf{uBGN}}& \multicolumn{1}{c}{\textbf{BFL}} \\ \hline
    \textbf{Normal} & \textbf{Cost} &       &       &       &       &  \\ \cline{2-7}
    \textbf{} & \textbf{} & 0.8    &0.147 & 0.201& & 0.180 \\
    \textbf{} & \textbf{} & -0.8  &  0.147 & 0.209 && 0.182 \\ \cline{2-7}
    \textbf{} & \textbf{QALYs} &       &       &       &       &  \\ \cline{2-7}
   \textbf{} & \textbf{} & 0.8   &  0.072 & 0.100 && 0.086 \\
    \textbf{} & \textbf{} & -0.8   &0.075 & 0.109 && 0.090 \\ \cline{2-7}
    \textbf{} & \textbf{INB} &       &       &       &       &  \\
\hline
    \end{tabular}%
  \label{rmse70N100_high}%
\end{table}%

\begin{table}[h]
  \centering
  \caption{RMSE for incremental Cost, QALYs and INB across scenarios with 70\% non-compliance, moderate correlation between outcomes and sample size $n=1000$. uBGN was not applied in settings with normal cost data. Numbers for INB have been rounded to the nearest integer.}
    \begin{tabular}{lcrrrrr} \hline
    \textbf{Cost distribution} & \textbf{} & \multicolumn{1}{c}{\textbf{$\rho$}}  & \multicolumn{1}{c}{\textbf{3sls (2sls)}} & \multicolumn{1}{c}{\textbf{uBN}}& \multicolumn{1}{c}{\textbf{uBGN}} & \multicolumn{1}{c}{\textbf{BFL}} \\ \hline
    \textbf{Normal} & \textbf{Cost} &       &       &       &       &  \\ \cline{2-7}
    \textbf{} & \textbf{} & 0.4   & 0.043 & 0.043 && 0.043 \\
    \textbf{} & \textbf{} & -0.4  & 0.043 & 0.043 && 0.043 \\ \cline{2-7}
    \textbf{} & \textbf{QALYs} &       &       &       &       &  \\ \cline{2-7}
    \textbf{} & \textbf{} & 0.4    & 0.022 & 0.022 & &0.022 \\
    \textbf{} & \textbf{} & -0.4   & 0.022 & 0.022 & &0.022 \\ \cline{2-7}
    \textbf{} & \textbf{INB} &       &       &       &       &  \\
\hline
    \end{tabular}%
  \label{rmse70N1000}%
\end{table}%

\begin{table}[h]
  \centering
  \caption{RMSE for incremental Cost, QALYs and INB across scenarios with 70\% non-compliance, high correlation between outcomes and sample size $n=1000$. uBGN was not applied in settings with normal cost data. Numbers for INB have been rounded to the nearest integer.}
    \begin{tabular}{lcrrrrr} \hline
    \textbf{Cost distribution} & \textbf{} & \multicolumn{1}{c}{\textbf{$\rho$}}  & \multicolumn{1}{c}{\textbf{3sls (2sls)}} & \multicolumn{1}{c}{\textbf{uBN}}& \multicolumn{1}{c}{\textbf{uBGN}} & \multicolumn{1}{c}{\textbf{BFL}} \\ \hline
    \textbf{Normal} & \textbf{Cost} &       &       &       &       &  \\ \cline{2-7}
    \textbf{} & \textbf{} & 0.8   & & 0.043 & &0.043 \\
    \textbf{} & \textbf{} & -0.8 &  0.043 & 0.043 & &0.043 \\ \cline{2-7}
    \textbf{} & \textbf{QALYs} &       &       &       &       &  \\ \cline{2-7}
    \textbf{} & \textbf{} & 0.8   &0.021  & 0.021 & &0.021 \\
    \textbf{} & \textbf{} & -0.8   &0.021 & 0.022 & &0.021 \\ \cline{2-7}
    \textbf{} & \textbf{INB} &       &       &       &       &  \\
\hline
    \end{tabular}%
  \label{rmse70N1000_high}%
\end{table}%

\clearpage
\subsection{Software code for performing  MI and 3sls in  Stata}
The code belows performs MI of total costs, total QALYs and the baseline EQ5D summary score \texttt{eq5d\_b}, by chained equations using predictive mean matching with the 5 nearest neighbours (specified by the option \texttt{knn(5)}. We include treatment received  \texttt{ txreceived} as a variable in the imputation models, and perform the imputations separately by randomised treatment arm, \texttt{rnd} (done by specifying the option \texttt{by (rnd) }.
  
After obtaining 50 imputations, 3sls estimates are obtained, and pooled following Rubin's rules.
\begin{small}
\begin{verbatim}
*MI
sort rnd 
mi set mlong
mi register imputed total_qalys total_costs eq5d_b  
mi impute chained (pmm, knn(5) ) total_qalys  eq5d_b total_costs=   txreceived,///
	 by (rnd) add(50)   rseed (110)

*3sls
mi estimate, cmdok: reg3 (total_costs total_qalys== txreceived), exog(rnd) ireg
mi estimate(inb: _b[total_qalys:txreceived] *30000 - _b[total_costs:txreceived]), ///
	cmdok: reg3 (total_costs total_qalys==txreceived), exog(rnd)

mi estimate (icer: _b[total_costs:txreceived]/_b[total_qalys:txreceived] ), ///
	cmdok: reg3 (total_costs total_qalys== txreceived), exog(rnd)
\end{verbatim}
\end{small}
\subsection{Software code for performing  MI and 3sls in R}
The following code performs MI by chain equations using PMM. The default option for the number of donors is 5, so we do not need to specify this. The code also  carries out 3sls using \texttt{systemfit} in R in each imputed data set. Rubin's rules for pooling the estimates across imputed sets are coded manually.

\begin{footnotesize}
\begin{verbatim}
rm(list=ls())
library(MASS,"AER",systemfit,mice,foreign)
data <- read.dta("../data.dta")

#the data must be ordered as follows total_qalys  total_costs   txreceived eq5d_b

#MI using MICE in R 
M<-50  #50 imputations
data<-data[order(data$rnd),]       # order data by arm

#we impute separate in each randomised arm
data0<-subset(data,rnd==0)
data1<-subset(data,rnd==1)
#the variable rnd must be dropped from data1 and data0

#rnd=1
imp1 <-mice(data1, m = M, method =c("pmm","pmm", "","pmm"),print=FALSE)
#rnd=0
imp0 <-mice(data0, m = M, method =c("pmm","pmm", "","pmm"),print=FALSE)

##merging data
mj<-imput_no<- rep(1:M, each=length(data$rnd))
data_mi <-data[rep(1:nrow(data), M), ]
data_mi$mj<-imput_no

data_mi$baseq_imput<-data_mi$qaly_imput<-<-array(NA, dim=M*length(data$rnd))
data_mi$cost_imput<-array(NA, dim=M*length(data$rnd))

for (j in 1:M){
  data_mi$baseq_imput[data_mi$mj==j]<-c(complete(imp0,j)$baseq , complete(imp1,j)$baseq)
  data_mi$qaly_imput[data_mi$mj==j]<-c(complete(imp0,j)$qaly  , complete(imp1,j)$qaly )
  data_mi$cost_imput[data_mi$mj==j]<-c(complete(imp0,j)$cost , complete(imp1,j)$cost)
}
######################################################
####  3sls 
######################################################
coef.c<-var.coef.c<-matrix(NA, nrow=M, ncol=1)
coef.q<-var.coef.q<-matrix(NA, nrow=M, ncol=1)
coef.inb<-var.coef.inb<-matrix(NA, nrow=M, ncol=1)
coef.icer.3sls<-matrix(NA, nrow=M, ncol=1)


for (j in 1:M){  
  eq.cost <-  data_mi$cost_imput[data_mi$mj==j] ~ 
					data_mi$exp[data_mi$mj==j]+ data_mi$baseq_imput[data_mi$mj==j]
  eq.qaly <-  data_mi$qaly_imput[data_mi$mj==j] ~
					 data_mi$exp[data_mi$mj==j]+ data_mi$baseq_imput[data_mi$mj==j]
  eq.Exp.Rand <- data_mi$exp[data_mi$mj==j] ~ 
					data_mi$rnd[data_mi$mj==j]+ data_mi$baseq_imput[data_mi$mj==j]
  
  System3eq <- list( cost=eq.cost, qaly=eq.qaly, trt.received = eq.Exp.Rand )
  reg3 <- systemfit(System3eq, method = "3SLS",
                    inst = ~ data_mi$rnd[data_mi$mj==j] + data_mi$baseq_imput[data_mi$mj==j])
  
  coef.q[j]<-reg3$coefficients["qaly_data_mi$exp[data_mi$mj == j]"]
  coef.c[j]<-reg3$coefficients["cost_data_mi$exp[data_mi$mj == j]"]  
  var.coef.q[j]<-diag(reg3$coefCov)["qaly_data_mi$exp[data_mi$mj == j]"]
  var.coef.c[j]<-diag(reg3$coefCov)["cost_data_mi$exp[data_mi$mj == j]"]
  cov<-reg3$coefCov["qaly_data_mi$exp[data_mi$mj == j]","cost_data_mi$exp[data_mi$mj == j]"]    
  coef.inb[j]<- coef.q[j]*30000-coef.c[j]
  var.coef.inb[j]<-30000^2*(var.coef.q[j])+  var.coef.c[j] -(2*30000*cov)  
  coef.icer.3sls[j]<-(coef.c[j])/(coef.q[j])  
 	}
###### Apply Rubin's rules ####
mean.c <-mean(coef.c)
var.c <-mean(var.coef.c)
with.var.c<- var.c   # within-imputation variance
bet.var.c<-(1/(M-1))*sum((coef.c-mean.c)^2)   
se.c<- sqrt(with.var.c+(1+1/M)*bet.var.c)

##Q
mean.q <-mean(coef.q)
var.q <-mean(var.coef.q)
with.var.q<- var.q   # within-imputation variance
bet.var.q<-(1/(M-1))*sum((coef.q-mean.q)^2)   
se.q<- sqrt(with.var.q+(1+1/M)*bet.var.q)

##INB
mean.inb <-mean(coef.inb)
var.inb <-mean(var.coef.inb)
with.var.inb<- var.inb   # within-imputation variance
bet.var.inb<-(1/(M-1))*sum((coef.inb-mean.inb)^2)   
se.inb<- sqrt(with.var.inb+(1+1/M)*bet.var.inb)

##ONLY Mean ICER
mean.icer <-mean(coef.icer.3sls)

res.3sls<-c(mean.c, se.c, mean.q, se.q, mean.inb, se.inb, mean.icer)

\end{verbatim}
\end{footnotesize}
\subsection{Model code for BFL in JAGS}
The following model can also be run in WinBUGs. This may be necessary if there is missing outcome data in the analysis at hand, since multivariate normal nodes cannot be partially observed in the current version of JAGS.

The vector \texttt{y=(txreceived, total\_qalys,total\_costs)}. The other variables are as before. 
 \begin{small}
\begin{verbatim}
model{
for (i in 1:n) {
   	  eq5d_b[i] ~  dnorm(qb, pr.qb)
  
    y[i,1:3] ~ dmnorm(mu[i,], pr[,])
    mu[i,1] <- b[1,1] + b[1,2] * rnd[i] 
     mu[i,2] <- b[2,1] + b[2,2]*b[1,2] * rnd[i] +  c*eq5d_b[i]  
     mu[i,3] <- b[3,1] + b[3,2]* b[1,2] *rnd[i] +  g*eq5d_b[i]            
			}
#		Priors:
pr[1:3,1:3]~dwish(R,3)
      R<-diag(1,3)

    for (k in 1:3) {
          b[k,1:2] ~ dmnorm(vm[],vp[,])
    }
    for (k in 1:2) {
          vm[k] <- 0
          vp[k,k] <- 0.1 
    }
    vp[1,2] <- 0
    vp[2,1] <- 0
                  pr.qb<- 1 / (qb.sd * qb.sd)
		qb.sd <- exp(log.qb.sd)
		log.qb.sd ~ dnorm(0, 0.01)
		qb ~ dunif(-0.5,1)
		c ~ dnorm(0, 0.01)
		g ~ dnorm(0, 0.01)
		
  #functionals of the parameters
  inb<- 30000*b[2,2]-b[3,2]
  icer<- b[3,2]/(b[2,2]+0.000001) #to avoid dividing by a zero

}
\end{verbatim}
\end{small}

\clearpage
\begin{small}

\end{small}

\end{document}